\RequirePackage{fix-cm}
\documentclass[smallcondensed]{svjour3}       
\smartqed  
\usepackage{graphicx}
\usepackage{xcolor}
\usepackage{xspace}
\usepackage{amssymb}
\usepackage{mathtools}
\usepackage{ulem}

\usepackage{float} 

\makeatletter
\def\moverlay{\mathpalette\mov@rlay}
\def\mov@rlay#1#2{\leavevmode\vtop{%
   \baselineskip\z@skip \lineskiplimit-\maxdimen
   \ialign{\hfil$\m@th#1##$\hfil\cr#2\crcr}}}
\newcommand{\charfusion}[3][\mathord]{
    #1{\ifx#1\mathop\vphantom{#2}\fi
        \mathpalette\mov@rlay{#2\cr#3}
      }
    \ifx#1\mathop\expandafter\displaylimits\fi}
\makeatother

\makeatletter
\newcommand{\xRrightarrow}[2][]{\ext@arrow 0359\Rrightarrowfill@{#1}{#2}}
\newcommand{\Rrightarrowfill@}{\arrowfill@\equiv\equiv\Rrightarrow}
\newcommand{\xLleftarrow}[2][]{\ext@arrow 3095\Lleftarrowfill@{#1}{#2}}
\newcommand{\Lleftarrowfill@}{\arrowfill@\Lleftarrow\equiv\equiv}
\makeatother

\usepackage{lineno}
\usepackage[inference]{semantic}
\usepackage{amsmath}

\usepackage{caption}
\usepackage{subcaption}

\usepackage{inconsolata}
\usepackage{verbatim}
\usepackage{url}
\usepackage[bookmarks=false]{hyperref}
\usepackage{fancyvrb}
\usepackage{todonotes}

\newcommand{\marioold}{{\sf\small MARIO}\xspace}
\newcommand{\mario}{{\sf\small MARIO$_{2}$}\xspace}
\newcommand{\ws}{{\sf\small WS}\xspace}
\newcommand{\ls}{{\sf\small LS}\xspace}

\newcommand{\policyone}{\textit{Policy 1}\xspace}
\newcommand{\policytwo}{\textit{Policy 2}\xspace}
\newcommand{\policythree}{\textit{Policy 3}\xspace}
\newcommand{\policyfour}{\textit{Policy 4}\xspace}

\begin{document}

\title{Declarative Application Management in the Fog\thanks{$^*$Author names are ordered alphabetically. All authors have contributed equally to this study. 
\\
\mbox{ \ }
\\
This work has been partly supported by project ``{\it Lightweight Self-adaptive Cloud-IoT Monitoring across Fed4FIRE+ Testbed}" (LiSCIo) funded by Fed4Fire+, by project ``\textit{Continuous QoS-compliant Management of Software Applications over the Cloud-IoT Continuum}" (CONTWARE) funded by the Conference of Italian University Rectors and by the Spanish Government (AEI) and the EU Commission (\textit{Fondo Europeo de Desarrollo Regional}) through grant number TIN2017-88547-P (MINECO / AEI / FEDER, UE).}
}
\subtitle{A bacteria-inspired decentralised approach}

\titlerunning{Declarative Application Management in the Fog}        

\author{Antonio Brogi$^*$       \and
        Stefano Forti$^{*}$          \and 
        \\Carlos Guerrero$^*$       \and
        Isaac Lera$^*$
}


\institute{A. Brogi \and S. Forti (\textit{corresponding author}) \at
              Department of Computer Science, University of Pisa \\
              \email{\{brogi, stefano.forti\}@di.unipi.it}           
           \and
           C. Guerrero \and I. Lera \at
           Department of Mathematics and Computer Science,  University of Balearic Islands\\
            \email{\{carlos.guerrero, isaac.lera\}@uib.es}
}

\date{Received: date / Accepted: date}

\maketitle

\begin{abstract}
Orchestrating next-gen applications over heterogeneous resources along the Cloud-IoT continuum calls for new strategies and tools to enable scalable and application-specific managements. Inspired by the self-organisation capabilities 
of bacteria colonies, we propose a declarative, fully decentralised application management solution, targeting pervasive opportunistic 
Cloud-IoT infrastructures. We present a customisable declarative implementation of the approach and validate its scalability through simulation over motivating scenarios, also considering end-user’s mobility and the possibility to enforce application-specific management policies for different (classes of) applications.
\end{abstract}

\section{Introduction}
\label{sec:intro}

In recent years, the problem of how to effectively manage distributed applications along the Cloud-IoT continuum~\cite{habibi2020fog,villari2019osmosis,pham2020survey,filali2020multi} -- i.e. in Fog computing settings -- has gained increasing attention from the scientific community~\cite{brogi2020place,applicationmanagementfogsurvey,VAQUERO201920}. Addressing such a problem is challenging due to the scale and heterogeneity that characterise Fog infrastructures, and due to the plethora requirements that next-gen IoT applications need to meet at run-time (e.g. end-to-end latency, success response rate, load-balancing)~\cite{Ghobaei2020}. 
In this context, new abstractions, methodologies and tools are needed to achieve two fundamental goals in Fog application management:

\begin{enumerate}
	\item[(\textit{i})] \textit{scalability} --- to master the provably exp-time complexity \cite{fogbrain} related to making informed decisions throughout the application lifecycle in large-scale settings, from first placement to undeployment, through scaling and migrations,
	\item[(\textit{ii})] \textit{flexibility} --- to enable application operators to specify \textit{ad-hoc} management policies capable of capturing a range of diverse application needs and to dynamically enforce such policies at runtime \cite{applicationmanagementfogsurvey}.
\end{enumerate}

\noindent
Most of the existing literature in the field of Fog application management has proposed centralised \textit{one-size-fits-all} solutions to make informed decisions, based on a global view of the available infrastructure~\cite{brogi2020place}. These solutions shine when dedicated computational resources are available, when optimisation metrics are fixed once for all, or when applications can be statically placed for long-enough periods --- thus amortising the costs of decision-making (e.g. in smart-home, gaming, virtual reality applications)~\cite{guerrero2018migration}. 
On the contrary, such solutions show their shortcomings when the available infrastructure is \textit{opportunistic}, built by exploiting a large number of heterogeneous (mobile) resource-constrained devices, or when distinct (classes of) applications running over the same infrastructure require optimising different metrics. Examples of these include disaster recovery or mass-event applications (e.g. concerts), which prevent employing centralised decision-making, demanding instead \textit{self-organisation capabilities} of application management~\cite{opportunisticfog2019,casadeiopportunistic,sarteco2019,dazzi2020}.

In our previous work~\cite{mario1}, we partly addressed the aforementioned limitations of centralised approaches by proposing a \textit{fully decentralised} and \textit{declarative} solution for \textsf{M}\textit{anaging} \textsf{A}\textit{pplications} \textsf{R}\textit{unning} \textsf{I}\textit{n} \textsf{O}\textit{pportunistic} Fog scenarios, and its open-source Prolog prototype \marioold. Being fully decentralised, \marioold is \textit{scalable}. Notably, it is also highly \textit{flexible}, since it enables application operators to declaratively specify \textit{ad hoc} application management policies, using a set of simple operations that trigger based on available monitored data. 
\marioold assumes application management agents to be in charge of single application instances. Besides, it assumes they can access monitored data on the nodes in their direct vicinity (viz. free hardware at neighbouring nodes), and on all nodes traversed by user requests targeting their application instance (viz. request rate, free hardware in traversed nodes, end-to-end path latencies).

In this article, pursuing the effort towards fully decentralised declarative application management in the Fog, we extend the preliminary modelling and results of~\cite{mario1}, and we open-source a new prototype\footnote{Available at \url{https://github.com/acsicuib/MARIO/tree/MarioII}.}, \mario. 
The most important novelties featured by \mario are:
\begin{description}
	\item[\textbf{Biological Parallel and Formal Operational Semantics} --] By establishing a parallel between our solution and the biological behaviour of colonies of bacteria, we refine the set of management operations that can be used to declare management policies, and we propose a formal operational semantics of such operations to formally characterise the emerging behaviour of \mario application management agents running over a Fog infrastructure.
	\item [\textbf{Limited Contextually Available Data} --] We reduce the amount of information available at application management agents, by limiting it to the free hardware at their host node and to the request rates from neighbouring nodes along with the end-to-end latency they experience. This dramatically reduces the information that node managers disclose to application management agents, placing \mario into more realistic Fog settings compared to \marioold.
	\item[\textbf{New Policies and Experiments with User Mobility} --] We propose four new management policies based on the new modelling and we extensively simulate them in YAFS~\cite{lera2019yafs} over a large-scale lifelike scenario, based on real data about taxi traces in Rome, Italy~\cite{rometaxis}. The obtained results confirm the scalability and flexibility of the new \mario, also proving it successful in managing applications in presence of end-users' mobility.
\end{description}

\noindent
Last, but not least, this article presents an extended discussion of related work to capture recent advances in the field of decentralised management solutions in Fog and Cloud-IoT settings.

\medskip
\noindent
The rest of this article is organised as follows. Section~\ref{sec:bioframework} describes the parallel among the emergent behaviour of bacteria colonies and our declarative fully-decentralised management solution, providing a formal operational semantics of the behaviour of \mario application management agents. Section~\ref{sec:prologprototype} describes a Prolog prototype implementation of such a solution and four declarative  (workload- and latency-aware) application management policies it supports. Those policies are showcased and thoroughly assessed via simulations over a lifelike scenario with user mobility in Section \ref{sec:experiments}. Finally, Section \ref{sec:related} discusses some related work and Section \ref{sec:conclusions} draws some concluding remarks and outlines some possible directions for future work.

\section{Modelling Bacteria-inspired Declarative Application Management}
\label{sec:bioframework}

\subsection{Biological Interpretation of \mario}

Seeking to bring the self-organisation capabilities of biological systems in opportunistic Fog application management, \mario strongly inspires from the behaviour of bacteria. Indeed, \mario relies upon fully-decentralised agents that perform three simple customisable operations, mimicking primitive biological functionalities.
In this section, we illustrate the correspondence between those simple operations and the emergent behaviour of bacteria, and we provide a formal operational semantics of the \mario decentralised management solution.

Application management agents, as bacteria, live in a \textit{host environment} represented by a Cloud-IoT infrastructure. Infrastructure nodes correspond to \textit{places} in the host environment where contextual deployment conditions (e.g. available hardware resources, client requests to the application, end-to-end application latencies, resource management policies) can be more or less favourable for the considered application, and consequently trigger different management operations. 
%
%
\noindent
Fig. \ref{fig:comparison} sketches the parallel between the considered biological functionalities of bacteria \cite{campbell} and the management operations featured by \mario agents. Particularly:

\begin{figure}
	\centering
	\includegraphics[width=0.85\textwidth, clip, trim=0.8cm 0cm 15cm 0cm]{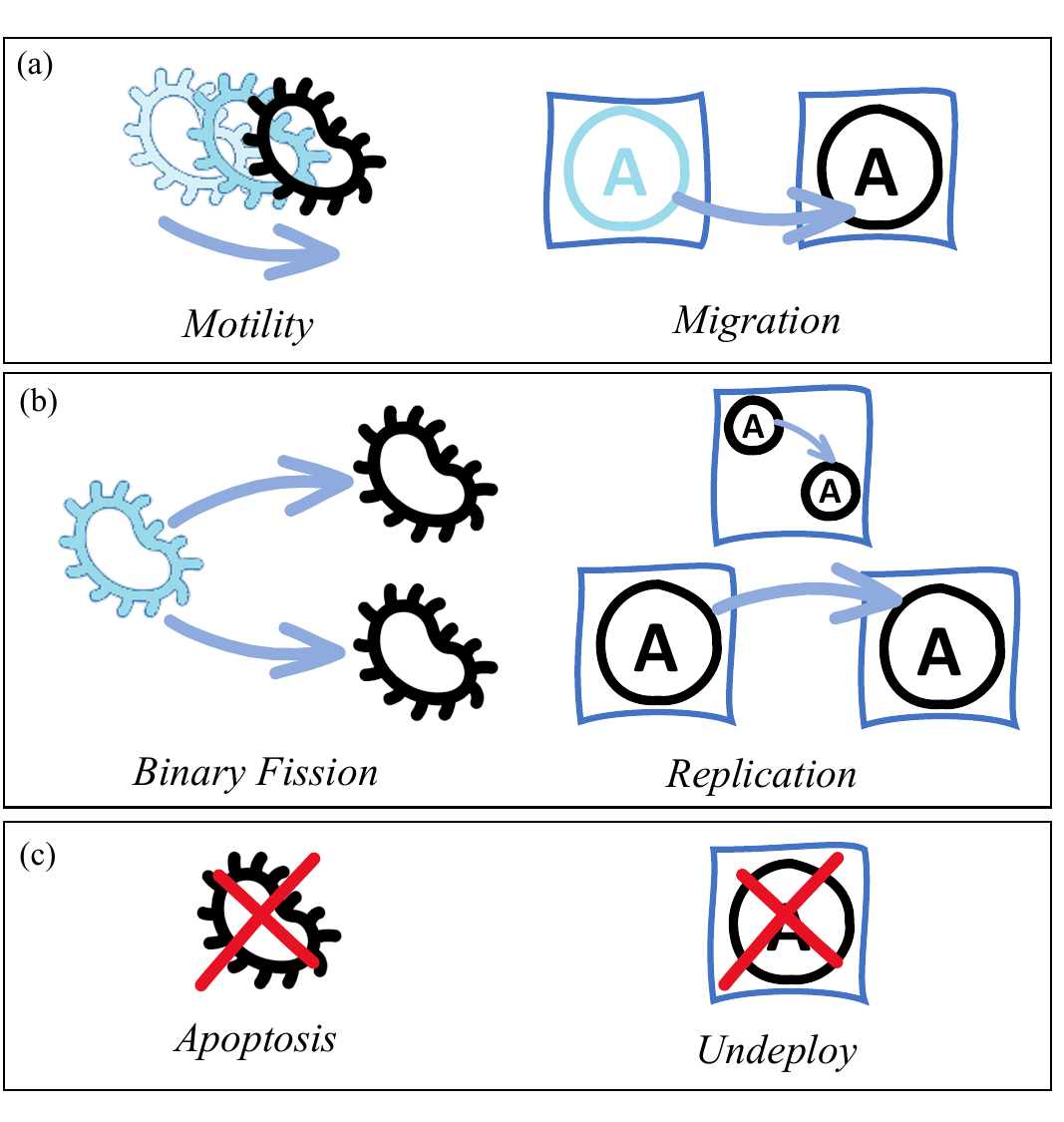}
	\caption{Bacteria functionalities and corresponding \mario operations.}
	\label{fig:comparison}
\end{figure}

\begin{description}
	\item[\textit{Motility}] Bacteria can move around by means of \textit{flagella}, i.e. lashes protruding from their bodies. Similarly, application instances in the \mario solution can perform \textit{migrations} from one deployment node to another according to their management policy and to the host environment around them. Fig. \ref{fig:comparison}(a) shows the analogy between bacteria mobility and application migration in \mario.
	\item[\textit{Binary Fission}] When their environment conditions are favourable, bacteria asexually reproduce via \textit{binary fission}, i.e. parent cells divide into two identical daughter cells. Similarly, application instances in the \mario solution can replicate into a new identical application instance, on either the same node or on a node in their vicinity\footnote{ Binary fission onto a neighbouring node enables application instances to replicate -- e.g. to handle increased workloads -- even when their current node lacks the needed resources. Note that such a behaviour could not be achieved by composing a replication onto the current node with a subsequent migration. },  depending on which one represents the more favourable place in their host environment according to the enforced application management policy. Fig.~\ref{fig:comparison}(b) shows the analogy between bacteria binary fission and application replication in \mario, when an application instance replicates in the same or in a different target node. 
	\item[\textit{Apoptosis}] Bacteria feature a form of programmed cell death called \textit{apoptosis}, i.e. bacteria can suicide when death of an individual translates into benefits for its colony \cite{bacteriaapoptosis}.  Similarly, application instances in \mario can decide to autonomically \textit{undeploy} from their host node, when needed according to their management policy. Fig.~\ref{fig:comparison}(c) shows the analogy between bacteria apoptosis and application instance undeployment in \mario.
\end{description}

\noindent
Last, but not least, bacteria are capable of \textit{sensing} their environment to properly react with some of the illustrated biological functionalities, when needed. Similarly, we assume that nodes feature decentralised \textit{monitoring} capabilities (e.g. those of FogMon \cite{fogmon2019,FORTI2020} or of the monitoring tools surveyed in \cite{taherizadeh2018monitoring}), enabling  application management agents to access information about their current deployment context. Pursuing our analogy, bacteria sensing capabilities correspond to the fact that application management agents can access information on the hardware resources at their current deployment node and on the amount of client requests they receive from each neighbouring node, along with the client-to-application latency such requests experience.

\subsection{Emerging Behaviour of \mario Bacteria}\label{sec:EnactmentAgentRequests}

We formally define here the emergent behaviour of \mario by means of a labelled transition system describing the evolution of application deployments over Cloud-IoT infrastructures.

\smallskip
\noindent 
Each application instance is associated with an \textit{application management agent} (i.e. a \textit{bacterion}), denoted by a tuple of the form
$\langle i, a, p \rangle$
where
\vspace{-0.15cm}
\begin{itemize}
	\item $i$ is the (unique)  identifier of the application management agent,
	\item $a$ is the managed application instance, and
	\item $p$ is the management policy associated to the application instance,  selecting one  management action to be executed, based on contextually available data.
\end{itemize}

\noindent 
\textit{Infrastructure nodes} are denoted by tuples of the form 
$\langle n, A_{n}, K_{n}, P_{n} \rangle$ 
where
\vspace{-0.15cm}
\begin{itemize}
	\item $n$ is a (unique) node identifier,
	\item $A_{n}$ is the set of application management agents running in node $n$, 
	\item $K_{n}$ is the contextual data available at node $n$ to the application management agents in $A_n$, and
	\item $P_{n}$ is (a predicate defining) the acceptance  policy employed by node $n$ to accept/reject deployment requests of application management agents.
\end{itemize}

\noindent 

An application management agent $b$ running on a node $n$ can decide to take different actions according to its management policy $p$ applied to the contextually available data $K_{n}$.
Namely an application management agent $b$ can request:
(1) to \textit{migrate} (together with its associated application instance) to a neighbour node $m$ of $n$,
(2) to \textit{replicate} itself in the same node $n$ where it is running, or in a neighbour node $m$ of $n$, by starting a new instance $b'$ of $b$, or 
(3) to \textit{undeploy} itself.
%

\smallskip
\noindent 
Rule (r1) describes the controlled evolution of a set of infrastructure nodes.
The set {\footnotesize ${\bf R}$} denotes the requests issued by the bacteria present in the nodes of the infrastructure as well as by human/autonomic application operators wishing to \textit{deploy} new application management agents.

Each request of an agent $b$ running in a node $n$ is assessed by (the node manager of) $n$ according to its acceptance management policy $P_n$. Requests to migrate or to replicate in a neighbour node $m$ of $n$ are also assessed by (the node manager of) $m$.
Accepted requests {\footnotesize ${\bf A}$} are all enacted in parallel, while non-accepted requests are inhibited and do not produce changes in the system configuration.

The set of accepted and inhibited requests labels the transition modelling the evolution of a set $N$ of infrastructure nodes. 
The effect of the transition is that the set of infrastructure nodes $N$ evolves into $N'$ by updating the set of bacteria running in each node $n \in N$ according to the
\textit{deploy}, \textit{migrate}, \textit{replicate} and \textit{undeploy} requests that were accepted.

\bigskip
\noindent
{\footnotesize
	$\begin{array}{ll}
	& 
	\begin{array}{l}
	{\bf R}
	\ =\
	\{
	\langle b, n, op \rangle
	\ \mid \
	\langle n, A_{n}, K_{n} , P_{n} \rangle \in N
	\ \wedge \
	b = \langle i, a, p \rangle \in A_n
	\ \wedge \
	op  = p(K_{n}) 
	\}
	\\
	\hspace{0.9cm} 
	\cup
	\ 
	\{ 
	\langle b, n, deploy \rangle
	\ \mid \
	\mbox{operator requests to deploy new agent $b$ in node $n$} 
	\}
	\\
	{\bf A} 
	\ = \
	\{x \mid x \in {\bf R} \ \wedge \ \forall n \in N: P_n(x)
	\}
	\ \subseteq \
	{\bf R}
	\\
	{\bf L} \ = \ {\bf A} \cup \{inhibited(w) \mid w \in {\bf R}-{\bf A} \}
	\end{array}
	\\
	(r1) & \cline{1-1}
	& 
	N \xrightarrow[]{{\bf L}} N'
	\end{array}$
}

\smallskip
\noindent
\textit{where} 

{\footnotesize
	\begin{tabbing}
		gggggggnatt\= \kill
		$N' =  \{  \langle n, A'_{n}, K_{n} , P_{n} \rangle \mid$
		\\
		\>$\langle n, A_{n}, K_{n} , P_{n} \rangle \in N  \ \wedge$
		\\
		\> 
		$A'_{n} = A_n \ $\=$ \cup \
		\{ b \ \mid \langle b, n, deploy \rangle \in {\bf A} \}$
		\\
		\>\>
		$\cup$ 
		$\{ b \ \mid \langle b, \_, migrate(n) \rangle \in {\bf A}
		\}$
		\\
		\>\>
		$\cup$ 
		$\{ b' \mid \langle b, \_, replicate(n) \rangle \in {\bf A} \wedge b = \langle i, a, p \rangle  
		\wedge b' = \langle j, a, p \rangle
		\}$ 
		\\
		\>\>
		$- \ \{ b \ \mid \langle b, n, undeploy \rangle \in {\bf A} \}$
		\\
		\>\>
		$- \ \{ b \ \mid \langle b, n, migrate(m) \rangle \in {\bf A} \}$
		\\
		$ \ \ \ \ \ \ \ \
		\}$
	\end{tabbing}
}

\bigskip
\noindent
Finally, the set of contextually available data at each node can (arbitrarily) change at any time.

\bigskip
\noindent
{\footnotesize
	$\begin{array}{ll}
	& 
	\begin{array}{l}
	\mbox{new data D available} 
	\wedge
	N' = \{ \langle n, A_{n}, K'_{n} , P_{n} \rangle | \langle n, A_{n}, K_{n} , P_{n} \rangle \in N \wedge 
	K'_{n}=D_{\mid n} \}
	\end{array}
	\\
	(r2) & \cline{1-1}
	& 
	N \xrightarrow[]{\tau} N'
	\end{array}$

	\smallskip
	\noindent
	\textit{where} 
	$D_{\mid n}$ denotes the set of data that become available to node $n$.
}




\section{Prototyping Bacteria-inspired Declarative Application Management}
\label{sec:prologprototype}

\noindent In this section,  after a short introduction to Prolog (Sect. \ref{sec:Prolog}),  we discuss how the bio-inspired fully-decentralised declarative management solution described in Sect. \ref{sec:bioframework} has been implemented in the \mario Prolog
prototype (Sect. \ref{sec:knowledgerep}). 
Then, we describe four lifelike, increasingly complex, management policies that can be declared and enforced in \mario as declarative Prolog programs (Sect. \ref{sec:policies}).

\subsection{Background: Prolog}
\label{sec:Prolog}
Prolog is a logic programming language based on first-order logic. 
A Prolog program is a finite set of \textit{clauses} of the form:
%
\begin{equation}\label{clause}
{\tt A :- B1, B2, ..., Bn.}
\end{equation}

\noindent
where {\tt A} is an atomic formula and where 
{\tt B1}, {\tt B2}, ..., {\tt Bn}  are (possibly negated) atomic formulas.
Clauses with empty premise are also called \textit{facts}.

The declarative reading of a clause like (\ref{clause}) is that {\tt A} is true if 
{\tt B1}, {\tt B2}, ..., {\tt Bn} are all true.
The procedural reading of a clause like (\ref{clause}) is that if {\tt B1}, {\tt B2}, ..., {\tt Bn} can be proven to hold (by SLD resolution), then {\tt A} is proven to hold.

By convention, variables begin with upper-case letters, lists are denoted by square brackets, and negation by {\tt\small \textbackslash +}.

ISO Prolog implementations\footnote{https://www.iso.org/standard/21413.html} allow to employ 
disjunctions (denoted by {\tt\small ;}) in the premise of clauses, as for instance in:
\begin{center}
	{\tt A :- (B;C),D.}
\end{center}
as a shorthand for:
\begin{center}
	{\tt A :- B,D.}
	\\
	{\tt A :- C,D.}
\end{center}

ISO Prolog implementations also feature set predicates like 
\begin{center}\texttt{findall(Template, Goal, Result)}\end{center} that can be exploited to find all successful solutions of \texttt{Goal} and collect the corresponding instantiations of \texttt{Template} in the list \texttt{Result}. If \texttt{Goal} has no solutions then \texttt{Result} is instantiated to the empty list.

\subsection{Knowledge Representation in Prolog}
\label{sec:knowledgerep}

\subsubsection{Available Contextual Data}
In this section, we describe the Prolog facts that  application management agents can access to make informed decisions based on the requirements of the application that they are managing, on the status of their deployment node, and on the  requests that their application instance is currently receiving. 

Application management agents can access applications descriptors, detailing the requirements and capabilities of single-service applications denoted as in

\begin{Verbatim}[fontfamily=zi4, fontsize=\small, frame=single, framesep=1mm, framerule=0.1pt, rulecolor=\color{gray}]
service(ServiceId, RequiredHW, MaxRequestRate, MaxLatencyToClient).
\end{Verbatim}

\noindent where {\small\tt ServiceId} is a unique application identifier, {\small\tt RequiredHW} is the amount of hardware resources required by the application\footnote{For the sake of simplicity, we represent generic hardware units as integers, as in \cite{GuerreroLJ19,mario1}. Accounting for different types of resources (e.g. RAM, CPU, HDD) is a straightforward extension.}, {\small\tt MaxRequestRate} is the maximum amount of client requests that a single application instance can handle, and {\small\tt MaxLatencyToClient} is the maximum end-to-end latency that incoming requests can tolerate.

Application management agents can also access information on the capabilities of their current deployment node denoted as in

\begin{Verbatim}[fontfamily=zi4, fontsize=\small, frame=single, framesep=1mm, framerule=0.1pt, rulecolor=\color{gray}]
node(NodeId, AvailableHW).
\end{Verbatim}

\noindent where \texttt{\small NodeId} is a unique node identifier within the network (e.g. IP address) and \texttt{\small AvailableHW} is the free hardware that the node currently features.

Running application instances, managed by application management agents in \mario, are denoted as in

\begin{Verbatim}[fontfamily=zi4, fontsize=\small, frame=single, framesep=1mm, framerule=0.1pt, rulecolor=\color{gray}]
serviceInstance(ServiceInstanceId, ServiceId, Node).
\end{Verbatim}

\noindent
where \texttt{\small ServiceInstanceId} is a unique service instance identifier, \texttt{\small ServiceId} is the identifier of the application descriptor, and \texttt{\small Node} is the node where the instance is currently deployed. 

Finally, application management agents can access data on the incoming requests targeting their application instance from neighbouring nodes, denoted as in

\begin{Verbatim}[fontfamily=zi4, fontsize=\small, frame=single, framesep=1mm, framerule=0.1pt, rulecolor=\color{gray}]
requests(ServiceInstanceId, Neighbour, RequestRate, LatencyToClient).
\end{Verbatim}

\noindent
where \texttt{\small ServiceInstanceId} is the unique identifier of the managed application instance, \texttt{\small Neighbour} is the identifier of the last-hop neighbouring node from which requests reach the application instance, \texttt{\small RequestRate} is the average number of requests targeting the application instance in the last time period, and \texttt{\small LatencyToClient} is the end-to-end latency associated with the path traversed by the incoming requests, from the client to the considered application instance.

\begin{figure}[h]
	\centering
	\begin{subfigure}[b]{0.49\textwidth}
		\centering
		\begin{Verbatim}[fontfamily=zi4, numbers=left, numbersep=5pt, fontsize=\small, numberblanklines=false, frame=single, framesep=1mm, framerule=0.1pt, rulecolor=\color{gray},  firstnumber=1,tabsize=2]
		service(videoBroadcast, 4, 10, 25).
		node(n42, 10).
		serviceInstance(s1, videoBroadcast, n42).
		requests(s1, self, 5, 0).
		requests(s1, n41,  1, 15).
		requests(s1, n41,  1, 16).
		\end{Verbatim}
		\caption{Prolog facts at \texttt{\small s1}}
		\label{fig:knowledge_base_prolog}
	\end{subfigure}
	\hfill
	\begin{subfigure}[b]{0.49\textwidth}
		\centering
		\includegraphics[width=\textwidth,clip, trim=0.8cm 11.5cm 16.5cm 0cm]{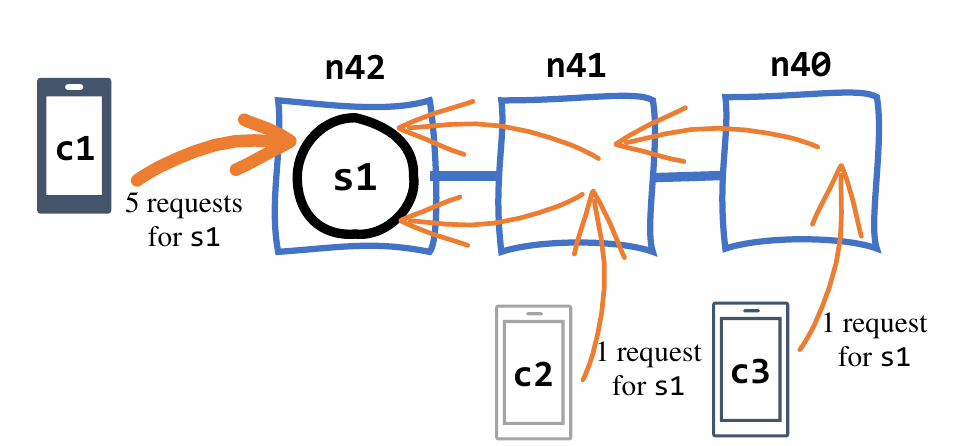}
		\caption{Requests to \texttt{\small s1} }
		\label{fig:knowledge_base_sketch}
	\end{subfigure}
	\caption{Knowledge base of a \mario application management agent.}
	\label{fig:knowledge_base}
\end{figure}

Facts of type {\tt\small service/4}, {\tt\small node/2}, {\tt\small serviceInstance/3} and {\tt\small requests/4} constitute the knowledge base accessible to each application management agent\footnote{ More preciely, with reference to the semantics presented in Sect. \ref{sec:EnactmentAgentRequests}, these Prolog facts implement the information on node idenfiers ($n$), sets of management agents ($A_n$), and available contextual data ($K_n$). The policies ($p$) of bacteria will be defined by the Prolog rules described in Sect.\ref{sec:policies}. }.
Fig. \ref{fig:knowledge_base_prolog} shows an example of knowledge base in which a video broadcasting service instance {\tt\small s1} (requiring 4 hardware units, serving at most 10 requests and tolerating an end-to-end latency of 25~ms) is running on node {\tt\small n42} featuring 10 free hardware units. 

\noindent
As sketched in Fig. \ref{fig:knowledge_base_sketch}, the application instance {\tt\small s1} is receiving 2 requests from application clients {\tt\small c2} and {\tt\small c3}, following two alternative routing paths (i.e. {\tt\small n41}-- {\tt\small n42} and {\tt\small n40}-- {\tt\small n41}-- {\tt\small n42}), respectively. As illustrated in Fig. \ref{fig:knowledge_base}, the application management agent is only aware of last hop information about incoming requests, i.e. it ignores that requests from \texttt{\small n41} come from different routing paths. Requests from clients in same node as the service instance are assumed to have null latency and indicated by Prolog facts such as the one at line 4 in Fig. \ref{fig:knowledge_base_prolog}, denoting direct requests of client {\tt\small c1} to  {\tt\small s1}.

\subsubsection{Management Operations}


The triggering of management operations described in Sect.\ref{sec:bioframework} (viz. undeploy, migrate, replicate) can be declaratively specified by means of simple Prolog rules. 
%

Each management operation and the associated {\tt\small TriggeringCondition} are declared as a Prolog rule of the form:

\begin{Verbatim}[fontfamily=zi4, fontsize=\small, frame=single, framesep=1mm, framerule=0.1pt, rulecolor=\color{gray}]
operation(OperationId, ServiceInstanceId, TargetNode) :- TriggeringCondition.
\end{Verbatim}

\noindent where {\tt\small OperationId} $\in \{${\tt\small undeploy}, {\tt\small migrate}, {\tt\small replicate}$\}$, {\tt\small ServiceInstanceId} is the service instance to which the management rule applies, and {\tt\small TargetNode} is the node targeted by the operation (set to {\tt\small self} when the operation targets only the current deployment node). 
%

Note that we exploit Prolog's clause ordering to express (simple) priorities among management operations (defined by predicate {\tt\small operation/2}).
%
%

%


Finally, we assume that application management agents can be aware of which operations that they requested during the last management cycles were inhibited by \mario.
This is done by including in their knowledge base facts denoted as in

\begin{Verbatim}[fontfamily=zi4, fontsize=\small, frame=single, framesep=1mm, framerule=0.1pt, rulecolor=\color{gray}]
inhibited(OperationId, ServiceInstanceId, TargetNode).
\end{Verbatim}

\subsection{Declaring Management Policies}
\label{sec:policies}

By means of four examples, we epitomise how it is possible to compactly declare management policies exploiting the ingredients described up to now in this section. Before starting, it is worth noting that management policies in \mario are actual Prolog programs, which can be written -- and debugged \cite{prologdebug} -- as such. Since Prolog relies upon a declarative programming paradigm, the management policies we will illustrate express their desired goals instead of the commands or steps that must be performed to reach out those.

\paragraph{Policy 1 (\textit{Workload-aware}) --} This policy is primarily intended to handle changes in the application workload by triggering suitable undeployment, migration and scaling operations. The main code\footnote{All utility predicates not explicitly listed in this article are available at: \url{https://github.com/acsicuib/MARIO/blob/MarioII/multi-agent-policies/scenarios/TaxiRome/policy/lib.pl}.} for \textit{Policy 1} is listed in Fig. \ref{fig:policy1_code}.

\begin{figure}[!h]
	\centering
	\begin{Verbatim}[fontfamily=zi4, numbers=left, numbersep=5pt, fontsize=\scriptsize, numberblanklines=false, frame=single, framesep=1mm, framerule=0.1pt, rulecolor=\color{gray},  firstnumber=1,tabsize=2]
	operation(undeploy,Si,self) :- 
	\+ requests(Si,_,_,_).
	
	operation(migrate,Si,M) :- 
	findall((H,R,L),requests(Si,H,R,L),Requests),
	serviceInstance(Si, S, _), service(S,_,MaxRequestRate,_),
	sumRequestRates(Requests,TotalRequestRate), TotalRequestRate>MaxRequestRate,    %(1a) 
	findall(K,member((K,_,_),Requests),Ms), sort(Ms,[M]), dif(M,self).              %(2)
	
	operation(replicate,Si,M) :- 
	findall((H,R,L),requests(Si,H,R,L),Requests),
	serviceInstance(Si, S, N), service(S,RequiredHW,MaxRequestRate,_),
	sumRequestRates(Requests,TotalRequestRate), TotalRequestRate>MaxRequestRate,    %(1a)
	mostRequestsFrom(Requests, M), 
	(
	(dif(M,self)
	; 
	(M=self, node(N,AvailableHW,_), AvailableHW >= RequiredHW))
	).
	\end{Verbatim}
	\caption{Policy 1: workload-aware management.}
	\label{fig:policy1_code}
\end{figure}

\noindent
First, an {\tt\small undeploy} operation is triggered when application instance {\tt\small Si} is currently not receiving any client requests, i.e. when the knowledge base of the considered application management agent does not ({\tt\small \textbackslash +} denotes Prolog negation) contain any {\tt\small requests/4} fact (lines 1--2).

\noindent
Triggering a {\tt\small migrate} operation requires to first collect all requests targeting application instance {\tt\small Si} along with their last hop {\tt\small H} and experienced latency {\tt\small L} (line 4)\footnote{Note that the call \texttt{\small findall((H,R,L),requests(Si,H,R,L),Requests)} in Fig. \ref{fig:policy1_code} (line 4) corresponds to retrieving the set \texttt{\small Request} of triples \texttt{\small (H,R,L)} such that  \texttt{\small requests(Si,H,R,L) holds.}}. A migration of instance {\tt\small Si} to node {\tt\small M} triggers when:
\begin{itemize}
	\item[(1a)] {\tt\small Si} cannot handle the total amount of incoming requests as it exceeds the {\tt\small MaxRequestRate} it can handle (line 6), \textbf{AND}
	\item[(2)] all the requests for {\tt\small Si} have the same neighbour {\tt\small M} (different from {\tt\small self}, viz. {\tt\small dif(M, self)}) as last hop in their routing path (line 7).
\end{itemize}

\noindent
Finally, a {\tt\small replicate} operation of {\tt\small Si} in node {\tt\small M} triggers when the aforementioned condition (1a)  holds. The chosen target node {\tt\small M} is the node from which most requests are arriving to {\tt\small Si} (line 12), and it can be one among the current deployment node ({\tt\small self}) or its neighbours. It is worth noting that if {\tt\small self} is the replication target, it must also feature enough hardware resources to host a new instance of the considered application (line 16). 

\paragraph{Policy 2 (\textit{Latency-aware}) --} This policy is primarily intended to reduce the client-to-application end-to-end latency to keep it below the maximum value tolerated by the application. The main code for {\textit Policy 2}  is listed in Fig. \ref{fig:policy2_code}.

\begin{figure}[!h]
	\centering
	\begin{Verbatim}[fontfamily=zi4, numbers=left, numbersep=5pt, fontsize=\scriptsize, numberblanklines=false, frame=single, framesep=1mm, framerule=0.1pt, rulecolor=\color{gray},  firstnumber=1,tabsize=2]
	operation(undeploy,Si,self) :- 
	\+ requests(Si,_,_,_).
	
	operation(migrate,Si,M) :- 
	findall((H,R,L),requests(Si,H,R,L),Requests),
	serviceInstance(Si, S, _), service(S,_,_,MaxLatencyToClient),
	member((_,_,Lat),Requests), Lat>MaxLatencyToClient,                 % (1b)
	findall(K,member((K,_,_),Requests),Ms), sort(Ms,[M]), dif(M,self).  % (2)  
	
	operation(replicate,Si,M) :- 
	findall((H,R,L),requests(Si,H,R,L),Requests),
	serviceInstance(Si, S, _), service(S,_,_,MaxLatencyToClient),
	member((_,_,Lat),Requests), Lat>MaxLatencyToClient,                 % (1b)
	findall((K,Rk,Lk),(member((K,Rk,Lk),Requests),Lk>MaxLatencyToClient),FilteredRequests),
	dif(FilteredRequests,[]), mostRequestsFrom(FilteredRequests, M).
	\end{Verbatim}
	\caption{Policy 2: latency-aware management.} 
	\label{fig:policy2_code}
\end{figure}

\noindent
The {\tt\small undeploy} operation triggers under the same condition of \textit{Policy 1}. Conversely, a {\tt\small migrate} operation of instance {\tt\small Si} to node {\tt\small M} triggers when:
\begin{itemize}
	\item[(1b)] {\tt\small Si} does not satisfy the maximum latency constraint for all requests arriving from {\tt\small M} (line 6), \textbf{AND}
	\item[(2)] all the requests for {\tt\small Si} have the same neighbour {\tt\small M} (different from {\tt\small self}) as last hop in their routing path (line 7).
\end{itemize}

\noindent
Finally, a {\tt\small replicate} operation of {\tt\small Si} in node {\tt\small M} triggers when the aforementioned condition (1b)  holds (line 11), and  {\tt\small M} is the neighbour of {\tt\small self} from which most requests to {\tt\small Si} not satisfying the maximum latency constraint are arriving (lines 12--13). 

\paragraph{Policy 3 (\textit{Workload-aware and Latency-aware}) --} This policy is intended to both handle changes in the application workload (as per \textit{Policy 1}) and to reduce the client-to-application end-to-end latency to keep it below the maximum value tolerated by the application (as per \textit{Policy 2}). The main code is listed in Fig. \ref{fig:policy3_code}.

\begin{figure}[!h]
	\centering
	\begin{Verbatim}[fontfamily=zi4, numbers=left, numbersep=5pt, fontsize=\scriptsize, numberblanklines=false, frame=single, framesep=1mm, framerule=0.1pt, rulecolor=\color{gray},  firstnumber=1,tabsize=2]
	operation(undeploy,Si,self) :- 
	\+ requests(Si,_,_,_).
	
	operation(migrate,Si,M) :- 
	findall((H,R,L),requests(Si,H,R,L),Requests),
	serviceInstance(Si, S, _), service(S,_,MaxRequestRate,MaxLatencyToClient),
	(
	(sumRequestRates(Requests,TotalRequestRate), TotalRequestRate>MaxRequestRate)      %(1a)
	; % OR
	(member((_,_,Lat),Requests), Lat>MaxLatencyToClient)                               %(1b)
	),
	findall(K,member((K,_,_),Requests),Ms), sort(Ms,[M]), dif(M,self).                  %(2)
	
	operation(replicate,Si,M) :- 
	findall((H,R,L),requests(Si,H,R,L),Requests),
	serviceInstance(Si, S, N), service(S,RequiredHW,MaxRequestRate,MaxLatencyToClient),
	(
	(sumRequestRates(Requests,TotalRequestRate), TotalRequestRate>MaxRequestRate)      %(1a)
	; % OR
	(member((_,_,Lat),Requests), Lat>MaxLatencyToClient)                               %(1b)
	),
	findall((K,Rk,Lk),(member((K,Rk,Lk),Requests),Lk>MaxLatencyToClient),FilteredRequests),
	(
	(FilteredRequests=[], mostRequestsFrom(Requests, M), 
	(
	(dif(M,self)
	; % OR
	(M=self, node(N,AvailableHW,_),AvailableHW>=RequiredHW))
	)
	)
	; % OR
	(dif(FilteredRequests,[]), mostRequestsFrom(FilteredRequests, M))
	).
	\end{Verbatim}
	\caption{Policy 3: workload- and latency-aware management.} 
	\label{fig:policy3_code}
\end{figure}

\noindent
The {\tt\small undeploy} still triggers under the same condition of \textit{Policy 1} and \textit{2}. With reference to properties (1a), (1b) and (2) described in \textit{Policy 1} and \textit{2}, a {\tt\small migrate} operation of instance {\tt\small Si} to node {\tt\small M} triggers when it holds ((1a) \textbf{OR} (1b)) \textbf{AND} (2). Particularly, when:
\begin{itemize}
	\item[(1a)] {\tt\small Si} cannot handle the total amount of incoming requests as it exceeds the {\tt\small MaxRequestRate} it can handle (line 7) \textbf{OR}
	\item[(1b)] {\tt\small Si} does not satisfy the maximum latency constraint for all requests arriving from {\tt\small M} (line 9),
	\\ \textbf{AND} (irrespectively of which between (1a) and (1b) holds, or if they both hold)
	\item[(2)] all the requests for {\tt\small Si} have the same neighbour {\tt\small M} (differing from {\tt\small self}) as last hop in their routing paths (line 11).
\end{itemize}

\noindent
Finally, a {\tt\small replicate} operation of {\tt\small Si} in node {\tt\small M} triggers when condition (1a) (line 16) \textbf{OR} (1b) (line 18)  hold. Then, in case (1a) holds and not (1b) (viz. the list of requests not satisfying latency constraints is empty), {\tt\small M} is chosen to be the node from which most requests are arriving to {\tt\small Si} (lines 20--22), and it can be one among the current deployment node ({\tt\small self}) or its neighbours (lines 24--26) as in \textit{Policy 1}. Otherwise, if (1b) holds, the target node {\tt\small M} is the neighbour of {\tt\small self} from which most requests to {\tt\small Si} not satisfying the maximum latency constraint are arriving (line 30) as in \textit{Policy 2}.

\paragraph{Policy 4 (\textit{Workload-aware and Latency-aware, with Memory}) --} This last policy, listed in Fig. \ref{fig:policy4_code}, has the same objectives as \textit{Policy 3}. However, it refines the behaviour of \textit{Policy 3} by avoiding to trigger a {\tt\small migrate} or a {\tt\small replicate} to a node {\tt\small M} if such operation has been inhibited by \mario (lines 4 and 14) during the last management cycles.
This simple refinement of \textit{Policy 3} allows the application management agent to find an alternative management operation to execute, thus avoiding to get stuck trying to request the same operation over and over, e.g. migrating/replicating instead to a node different from those that recently refused its requests.

\begin{figure}[!h]
	\centering
	\begin{Verbatim}[fontfamily=zi4, numbers=left, numbersep=5pt, fontsize=\scriptsize, numberblanklines=false, frame=single, framesep=1mm, framerule=0.1pt, rulecolor=\color{gray},  firstnumber=1,tabsize=2]
	operation(undeploy,Si,self) :- 
	\+ requests(Si,_,_,_).
	
	operation(migrate,Si,M) :- 
	\+ inhibited(migrate,Si,M),                                                          % memory
	findall((H,R,L),requests(Si,H,R,L),Requests),
	serviceInstance(Si, S, _), service(S,_,MaxRequestRate,MaxLatencyToClient),
	(
	(sumRequestRates(Requests,TotalRequestRate), TotalRequestRate>MaxRequestRate)      %(1a)
	; 
	(member((_,_,Lat),Requests), Lat>MaxLatencyToClient)                               %(1b)
	),
	findall(K,member((K,_,_),Requests),Ms), sort(Ms,[M]), dif(M,self).                  %(2)
	
	operation(replicate,Si,M) :- 
	findall((H,R,L),(requests(Si,H,R,L),\+inhibited(replicate,Si,H)),Requests),          % memory
	serviceInstance(Si, S, N), service(S,RequiredHW,MaxRequestRate,MaxLatencyToClient),
	(
	(sumRequestRates(Requests,TotalRequestRate), TotalRequestRate>MaxRequestRate)      %(1a)
	; 
	(member((_,_,Lat),Requests), Lat>MaxLatencyToClient)                               %(1b)
	),
	%
	findall((K,Rk,Lk),(member((K,Rk,Lk),Requests),Lk>MaxLatencyToClient),FilteredRequests),
	(
	(FilteredRequests=[], mostRequestsFrom(Requests, M), 
	(
	(dif(M,self)
	; 
	(M=self, node(N,AvailableHW,_),AvailableHW>=RequiredHW))
	)
	)
	; 
	(dif(FilteredRequests,[]), mostRequestsFrom(FilteredRequests, M))
	).
	\end{Verbatim}
	\caption{Policy 4: workload- and latency-aware management, with memory.} 
	\label{fig:policy4_code}
\end{figure}


\section{Experimental results}
\label{sec:experiments}

In this section, we first describe (Sect. \ref{sec:setup}) a lifelike use case that we devised to validate \textit{Policies 1}--\textit{4} presented in Sect. \ref{sec:policies}. 
Then (Sect. \ref{sec:results}), we present the results of 
extensive simulations of the proposed policies over such scenario.
The simulation were carried on with the YAFS simulator (Sect. \ref{sec:archic}).
%
Videos of the executed simulation are online\footnote{Available at: \url{https://github.com/acsicuib/MARIO/tree/MarioII/videos}} and show how the simulated \mario system evolves over time while reacting to user movements as per \textit{Policies 1}--\textit{4}.

\subsection{YAFS simulation
}\label{sec:archic}
All simulations were carried on with YAFS~\cite{lera2019yafs}\footnote{To allow replicating the experiments, all code needed to run them is available at: \url{https://github.com/acsicuib/MARIO/tree/MarioII}.},
a Python-based, lightweight and easy to configure, discrete event simulator (DES) for Fog computing scenarios.

Each application management agent is modelled by a DES process that runs in a specific node of the topology. Each DES agent also represents a unique service instance, and it gathers its local facts (node, service, serviceInstance and requests) from its knowledge base $K_n$ as shown in Fig.\ref{fig:knowledge_base}. The DES agent periodically calls  a Prolog execution to evaluate these facts with the rules assigned and it sends its requested operation (viz., a tuple $\langle b, n, op \rangle$ with op $\in$ \{undeploy, migrate, replicate\} defined in section~\ref{sec:EnactmentAgentRequests}) to \mario. 

\mario is in turn another DES process that periodically evaluates all operation requested by the DES agents. \mario chooses the (last) operation requested by each agent and it either performs or inhibits the required operation. 
When \mario finishes evaluating all the requests, the routing network process of linking requests between users and service instances is updated, and new service instances or new allocations are considered. The coherence of the model is ensured by the fact that each DES process is an atomic process inside the simulator. 

\subsection{Experiments setup}\label{sec:setup}

\begin{figure}[!h]
	\centering
	\includegraphics[width=0.9\textwidth]{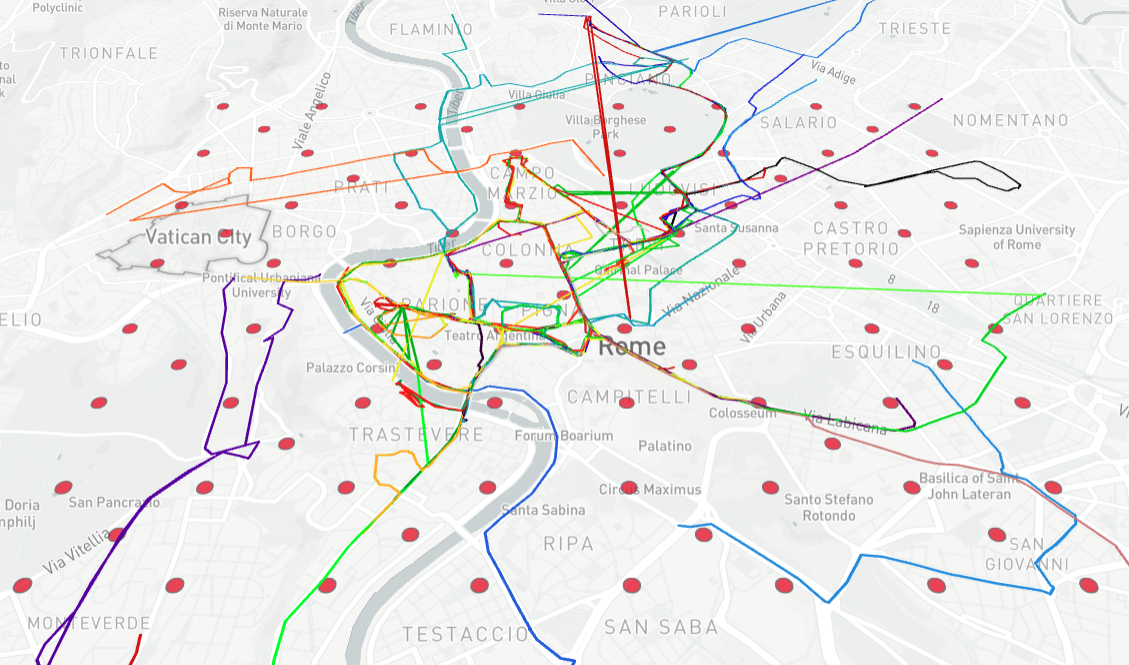}
	\caption{Taxi traces and Fog nodes projected on the map of Rome.}
	\label{fig:rome2d}
\end{figure}

To assess \textit{Policies 1}--\textit{4}, we have defined a lifelike scenario based on taxi mobility in the city of Rome, Italy. We have considered a region of Rome of 4 $\times$ 4 km$^{2}$, and we have simulated 51 users based on actual taxi traces taken from~\cite{rometaxis}. The considered area of Rome and the user mobility traces are sketched in Figure~\ref{fig:rome2d}. 

\begin{figure}[!h]
	\centering
	\includegraphics[width=0.5\textwidth]{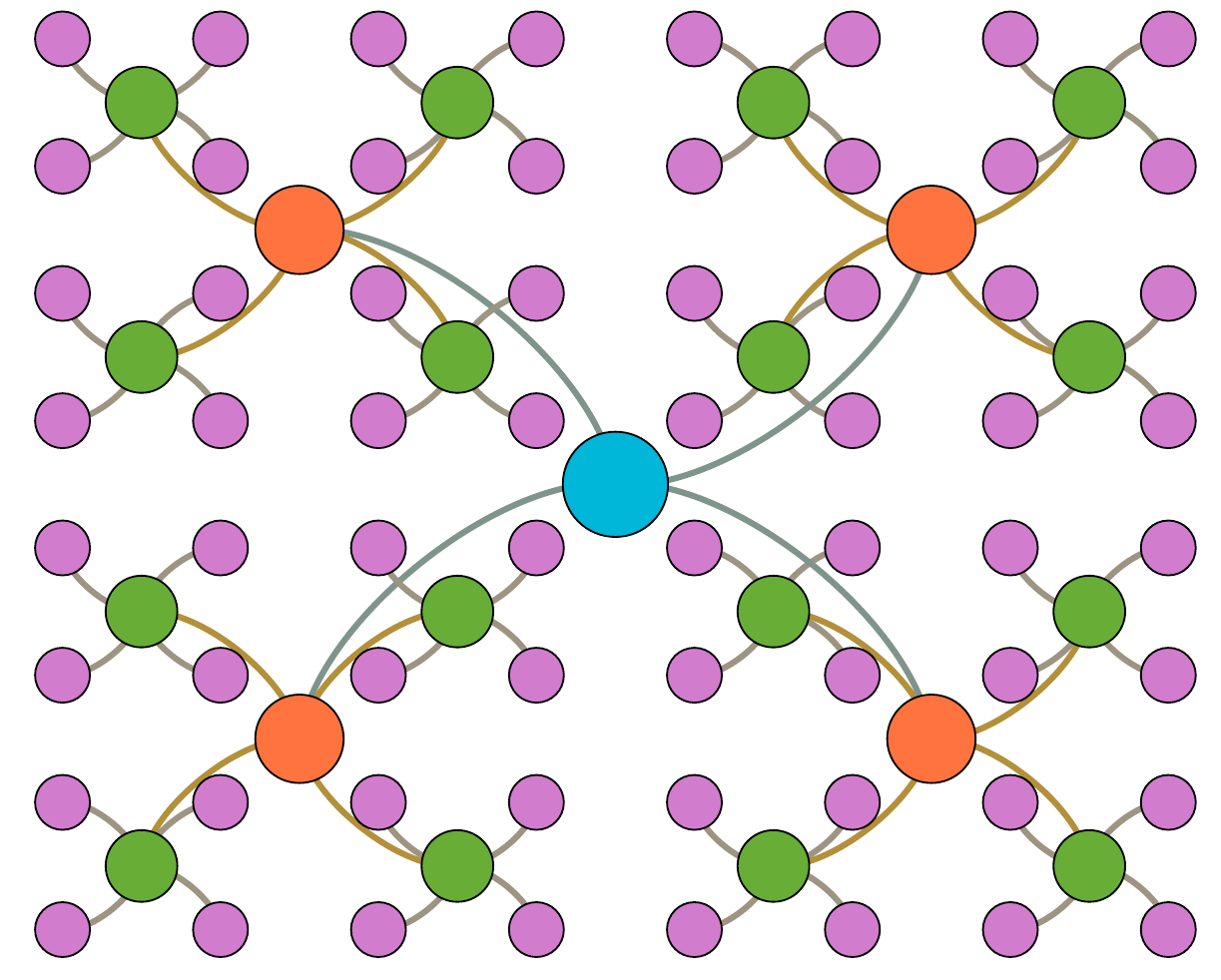}
	\caption{Infrastructure topology.}
	\label{fig:topo}
\end{figure}

Since, to the best of our knowledge, there is not publicly available data on the location of the cellular antennas in Rome, we uniformly distributed $64$ Access Points (APs) for the experiments using a grid distribution with AP-to-AP distance of 500 m. 
As shown in Fig.~\ref{fig:topo}, we consider a tree-based infrastructure with a total of $85$ nodes of heterogeneous capacity, where the APs are the edge nodes.
The nodes are distributed in a 4-tiered hierarchy consisting of:
\begin{itemize}
	\item 1 central Cloud (denoted by an azure circle) with virtually unbounded hardware and 3 ms link latency  
	to the nodes in the lower tier, 
	\item 4 mini-datacentres (orange circles) with 9 hardware units and 2 ms link latency  to the nodes in the lower tier, 
	\item 16 edge nodes (green circles) with 6 hardware units and 3 ms latency to the AP nodes, 
	and 
	\item 64 APs (purple circles) with 1 hardware unit.
\end{itemize}

\noindent Realistically, we assume that moving users always connect to the closest AP and we track 
the sequence of AP handovers (i.e. users changing their AP connection). These handovers are indeed responsible for the dynamic conditions of the experiments. Users are randomly split in two groups -- each group using different applications -- keeping the same average number of handovers in both groups. 
%


For the purposes of the experiments, we have considered six different applications declared as listed in Fig.~\ref{fig:appfeatures}. 
%
The six applications are divided into two types:
\begin{itemize}
	\item  \textit{workload-sensitive} (\ws, {\tt\small app1}--{\tt\small app3}), capable of handling a low request rate ($5$), and 
	\item \textit{latency-sensitive} applications (\ls, {\tt\small app4}--{\tt\small app6}), needing to have very low end-to-end latencies from their clients (viz. $1$ ms).
	
\end{itemize}

\begin{figure}[!h]
	\centering
	\begin{Verbatim}[fontfamily=zi4, fontsize=\small, frame=single, framesep=1mm, framerule=0.1pt, rulecolor=\color{gray}]
	% service(ServiceId, RequiredHW, MaxRequestRate, MaxLatencyToClient).
	
	%% workload-sensitive %%
	service(app1, 1, 5, 10).
	service(app2, 1, 5, 10).
	service(app3, 1, 5, 10).
	
	%% latency sensitive %%
	service(app4, 1, 40, 1).
	service(app5, 1, 40, 1).
	service(app6, 1, 40, 1).
	\end{Verbatim}
	\caption{Example applications.}
	\label{fig:appfeatures}
\end{figure}

Under the aforementioned conditions, \ws applications are expected to move closer to edge nodes when \policyone (workload-aware) and \policythree and \policyfour (workload- and latency-aware) are used, and to remain in the cloud when \policytwo (latency-aware) is used. 
On the other hand, \ls applications are expected to move closer to edge nodes when  \policytwo (latency-aware) and \policythree and \policyfour (workload- and latency-aware)  are used, and to remain in the cloud when \policyone (workload-aware) is used. 

\begin{figure}[!h]
	\centering
	\includegraphics[width=0.95\textwidth]{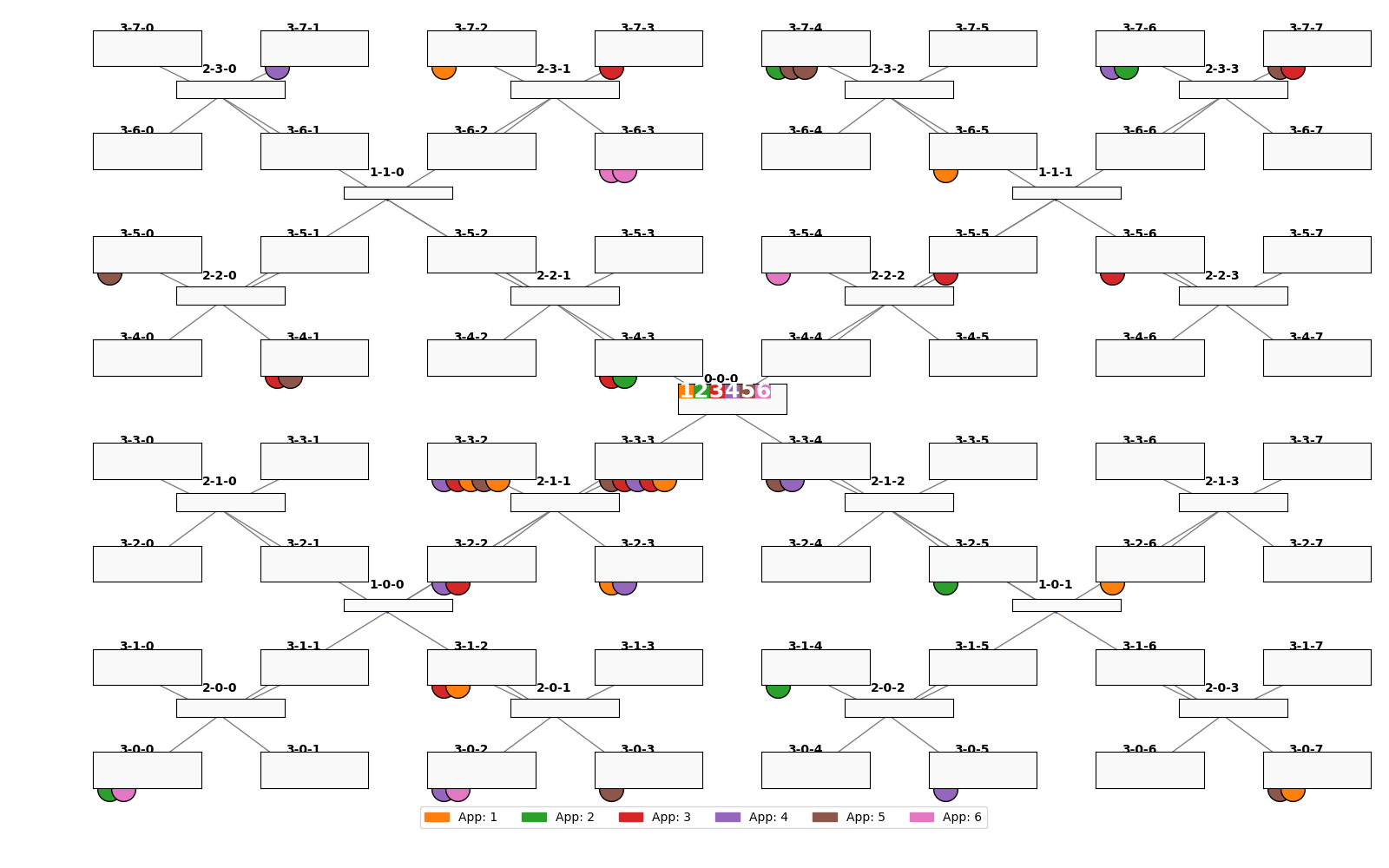}
	\caption{Initial experiment setup. 
	}
	\label{fig:usermovements}
\end{figure}

We assume that each user  requests only one application and that half of the users use \ws applications, and the other half use \ls applications.
We also take a uniform distribution of users' request rates, with each user sending a request every $40$ units of simulation time.
%
%
Each simulation is executed for 52,000 simulation units, because of the 25 user movements that are updated every 2,000 simulation units.
%
The simulator also makes the state of the scenario evolve by changing the AP associated to each users as the simulation goes forward, simulating a total of 81 AP handovers. 
Figure~\ref{fig:usermovements} shows the initial deployment of the applications (denoted by numbered squares in the central Cloud node) and the connections of the users to the APs (denoted by semicircles below the nodes). 

Four experiments were designed and carried out by assigning different management policies to different application types. In the initial state of the simulation, all the applications only have one instance deployed at the central Cloud node. 

\subsection{Results and discussion}\label{sec:results}

The objective of our simulations is to assess both dynamic (i.e. transient) and static (i.e. steady) application deployment conditions.
On one hand, they aim at assessing how promptly the proposed declarative, fully-decentralised management policies can  respond to application users' mobility, to workload variations and to violations of latency constraints required by running application instances.
On the other hand, they aim at assessing how the policies converge to a steady allocation of the applications over the available Fog infrastructure if  no changes occur for a sufficiently long period of time.


As mentioned before, changes in the considered scenario are produced by the movements of the users. 
Every time a user changes the AP it connects to:
\begin{enumerate}
	\item a number of {\tt\small migrate}, {\tt\small replicate} or {\tt\small undeploy} operations trigger, and 
	\item after a finite number of management cycles, if no other AP handover happens, the instances affected by users' movements do not require further actions, or they get rejected (viz. \textit{inhibited}) by the node manager to which they are directed.
\end{enumerate}

To illustrate such behaviour, we first present a set of plots where the number of current instances in the system and the type of operation selected by each of them is shown for each evaluation of the management rules\footnote{For the sake of the analysis, the plots also represent instances that do not require any action. These cases are represented in orange and named  {\tt\small nop}, to means that no action/operation is requested by a given service instance.}. Additionally, the number of user handovers between two evaluations of the management rules is also represented. From the analysis of these plots, we can observe the number of \mario executions that is needed after the user movements to achieve a steady allocation of the service instances. This is represented by the number of periods including {\tt\small migrate}, {\tt\small replicate} or {\tt\small undeploy} operations after a user movement. If there is no period without these types of operations between user movements, this means that application management agents are still \textit{searching} for a suitable allocation of their service instance, i.e. a temporarily steady allocation.

\begin{figure}[!h]
	\begin{subfigure}[b]{0.9\textwidth}
		\centering
		\includegraphics[width=0.8\textwidth]{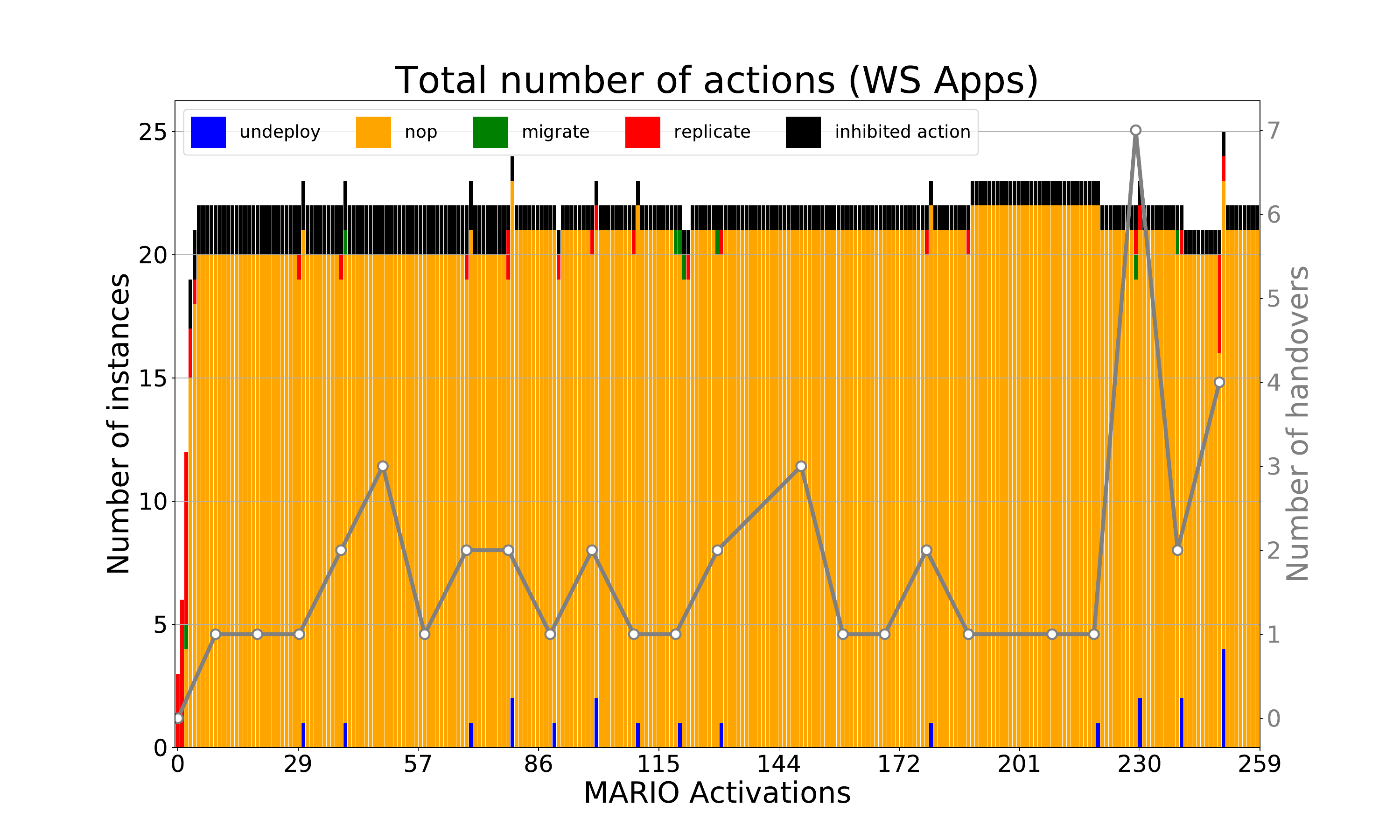}
		\caption{\ws applications}
		\label{fig:ap1b}
	\end{subfigure}
	\hfill
	\begin{subfigure}[b]{0.9\textwidth}
		\centering
		\includegraphics[width=0.8\textwidth]{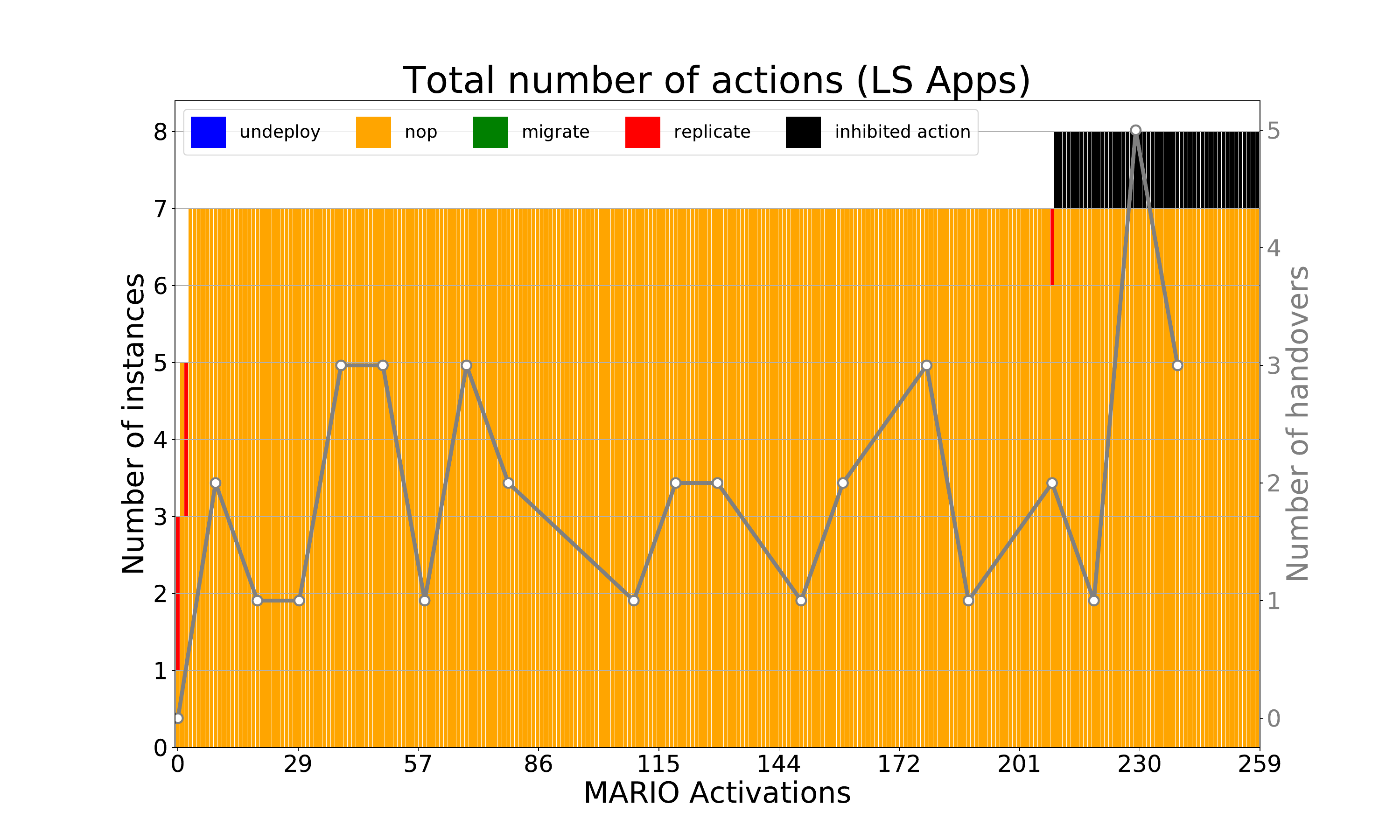}
		\caption{\ls applications}
		\label{fig:ap1c}
	\end{subfigure}
	\hfill    
	\begin{subfigure}[b]{0.9\textwidth}
		\centering
		\includegraphics[width=0.8\textwidth]{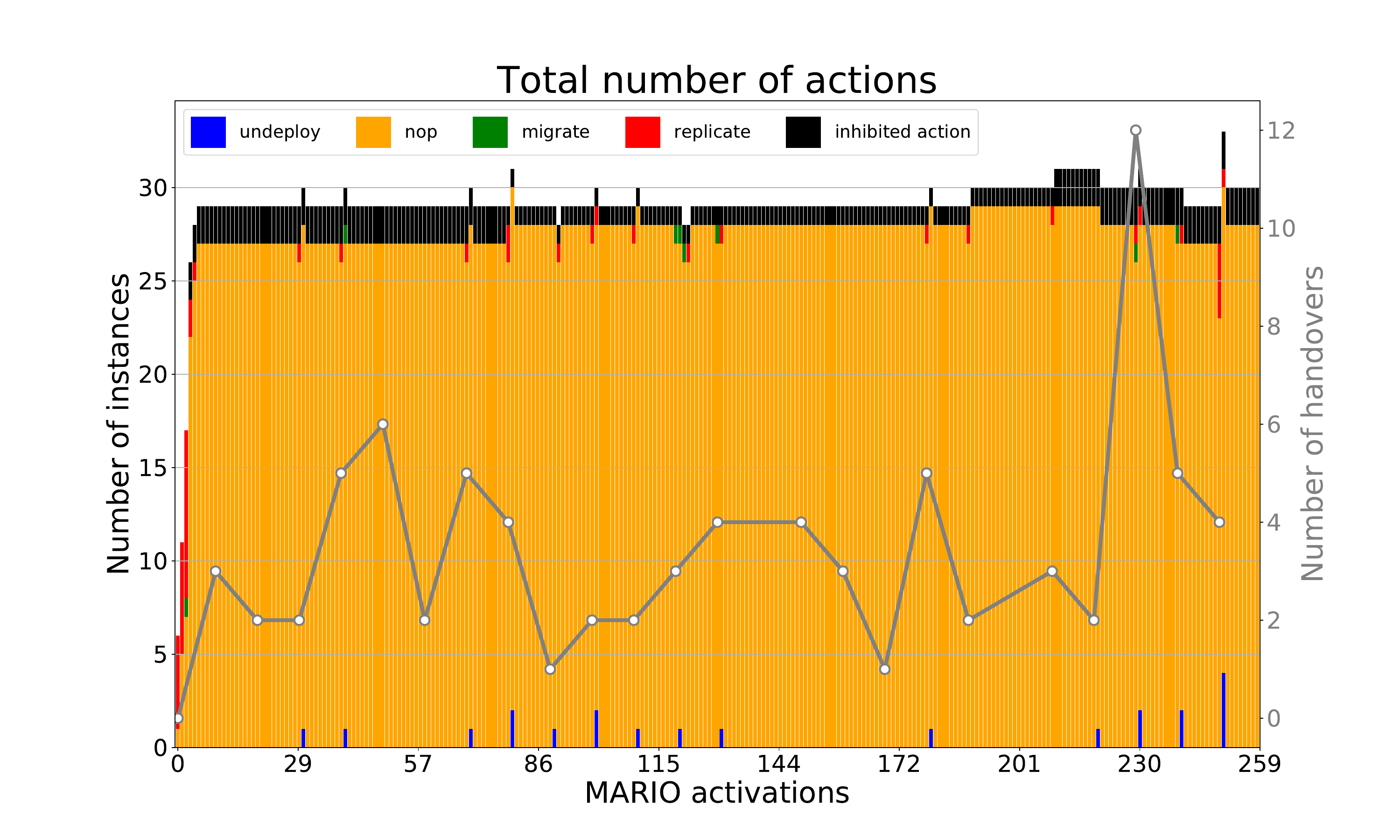}
		\caption{All applications}
		\label{fig:ap1a}
	\end{subfigure}
	\caption{Experiments with \policyone. }
	\label{fig:actionsP1}
\end{figure}

\paragraph{\policyone in Action --} The bar plots in Figure~\ref{fig:actionsP1} show the type of operations executed by the application instances along the simulation cycles, when \policyone is enforced by application management agents. Each bar is divided according to the number of actions of each type that has been triggered by running application managers. As \policyone is workload-aware, new instances of the application are generated when the already running instances are not able to handle all incoming requests. 

It is worth noting that \ws apps were defined with a low capacity of request response and, on the contrary, the \ls were defined with a low capacity of request response. From Figures~\ref{fig:ap1b} and~\ref{fig:ap1c}, we observe that the total number of application instances in the experiment settles at around 20-25 for the \ws applications (with all types of operation along the simulation) and at around 7 for the \ls applications with mainly only {\tt\small nop} and \textit{inhibited} operations (orange and black). 
The observed emerging behaviour of the \ws apps (Figure~\ref{fig:ap1b}) corresponds to the case in which the initial instances in the central Cloud cannot handle the total number of requests. Hence, they replicate and migrate closer to the users as per \policyone. On the contrary, the behaviour observed in the case of \ls apps (Figure~\ref{fig:ap1c}) corresponds to the case that the initial single instances of the applications are enough to handle all the request in the system (even though constantly violating the latency requirements). Consequently, the \ls applications do not replicate, neither migrate  closer to users as per \policyone.

From the plot that aggregates the operations from both sets of applications (Figure~\ref{fig:ap1a}), although some applications instances are continuously blocked (corresponding to the \textit{inhibited} operations), the system achieves a steady allocation after each set of user movements (represented with the dots of the blue line plot). Indeed, user movements are followed by a small number of management cycles where {\tt\small undeploy}, {\tt\small migrate}, and {\tt\small replicate} operations take place. The cycles needed to reach a temporarily steady condition settle on average between 1 and 2, which translates into the ability of \mario to promptly react to user movements. The \textit{inhibited} actions indicate that, although the allocation is stable, there are some applications that are not yet running in suitable nodes. They are trying to execute {\tt\small migrate} or {\tt\small replicate} operations but the node manager they are demanding to perform the operations rejects their request (
because the node does not feature sufficient free hardware to host them).

\begin{figure}[]
	\begin{subfigure}[b]{0.9\textwidth}
		\centering
		\includegraphics[width=0.8\textwidth]{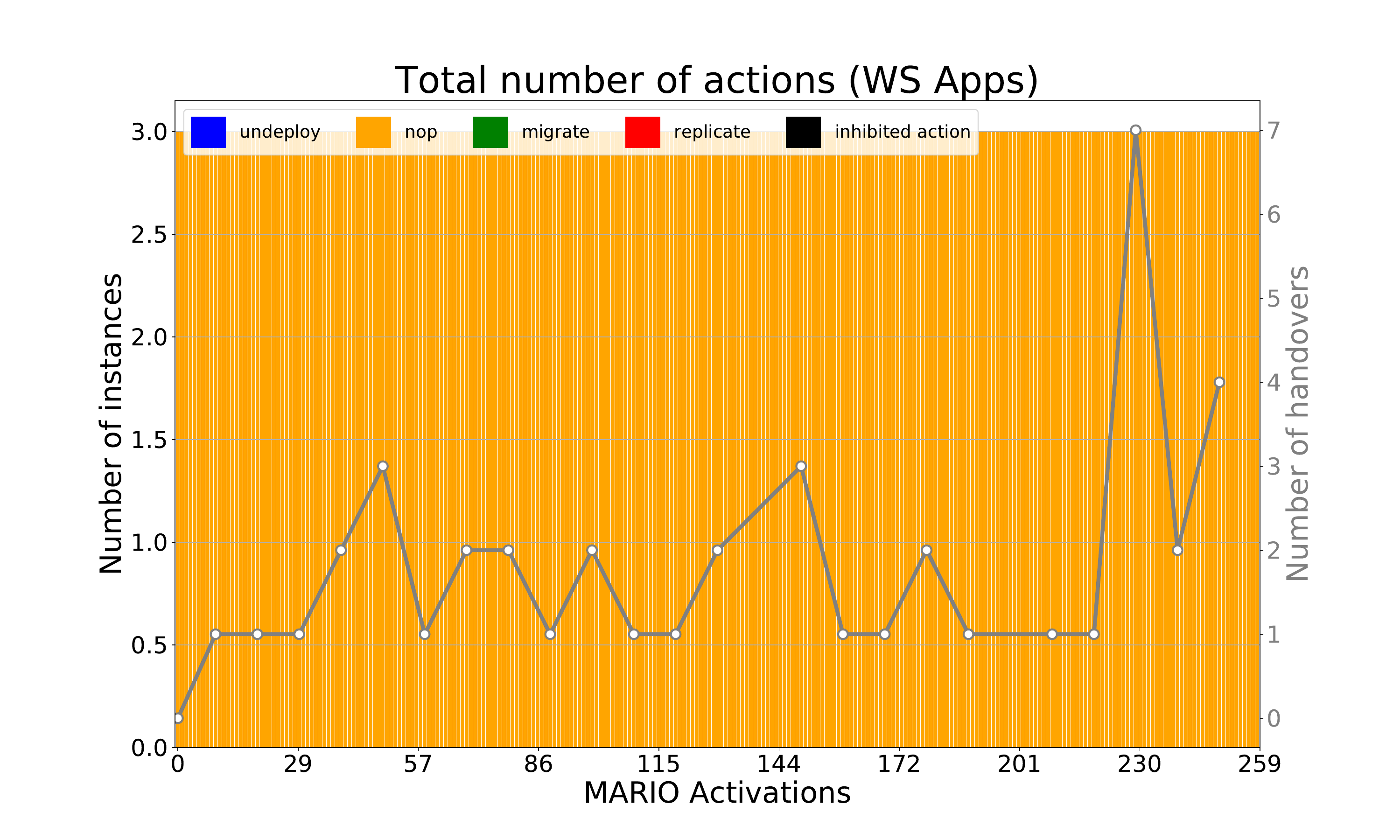}
		\caption{\ws application
			s}
		\label{fig:ap2b}
	\end{subfigure}
	\hfill
	\begin{subfigure}[b]{0.9\textwidth}
		\centering
		\includegraphics[width=0.8\textwidth]{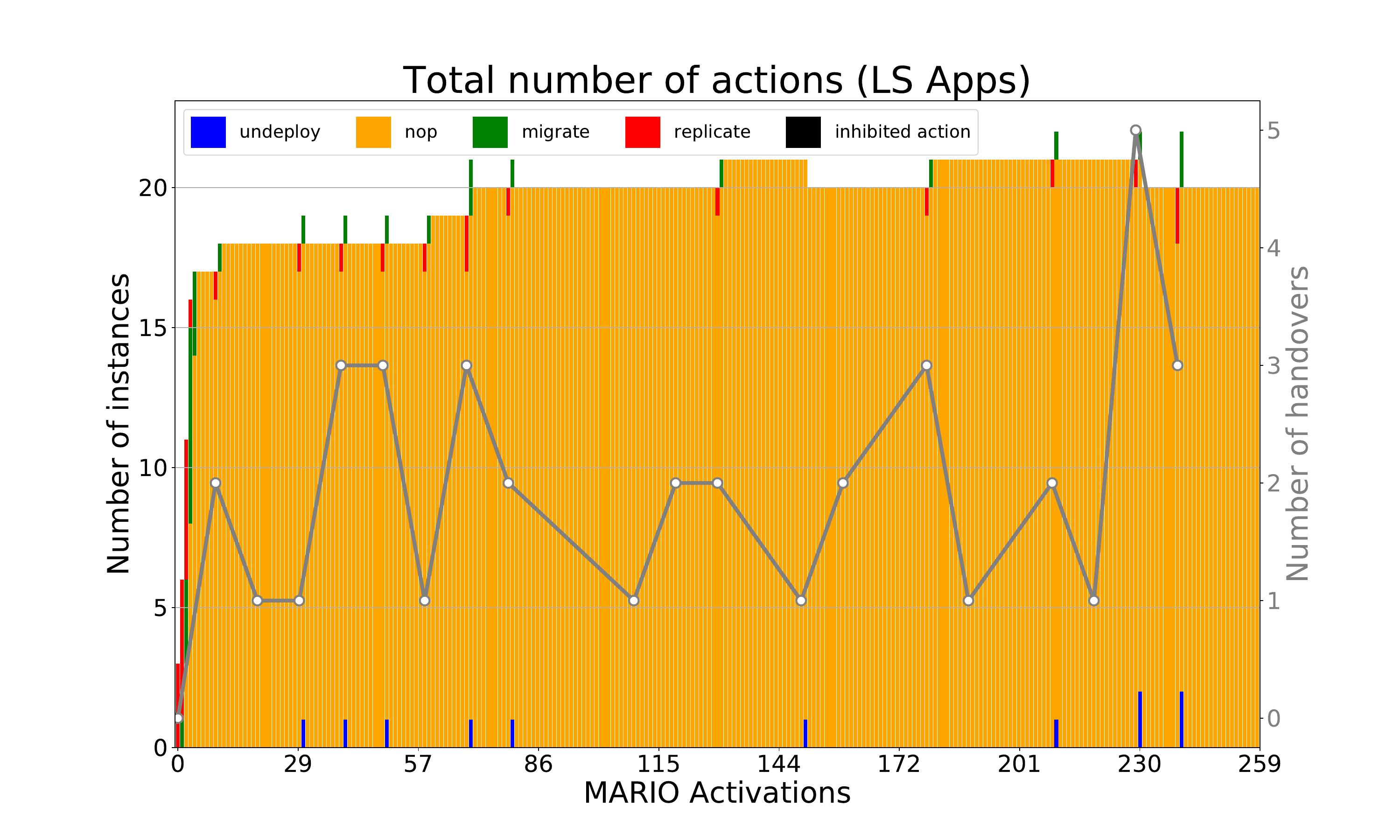}
		\caption{\ls applications}
		\label{fig:ap2c}
	\end{subfigure}
	\hfill
	\centering
	\begin{subfigure}[b]{0.9\textwidth}
		\centering
		\includegraphics[width=0.8\textwidth]{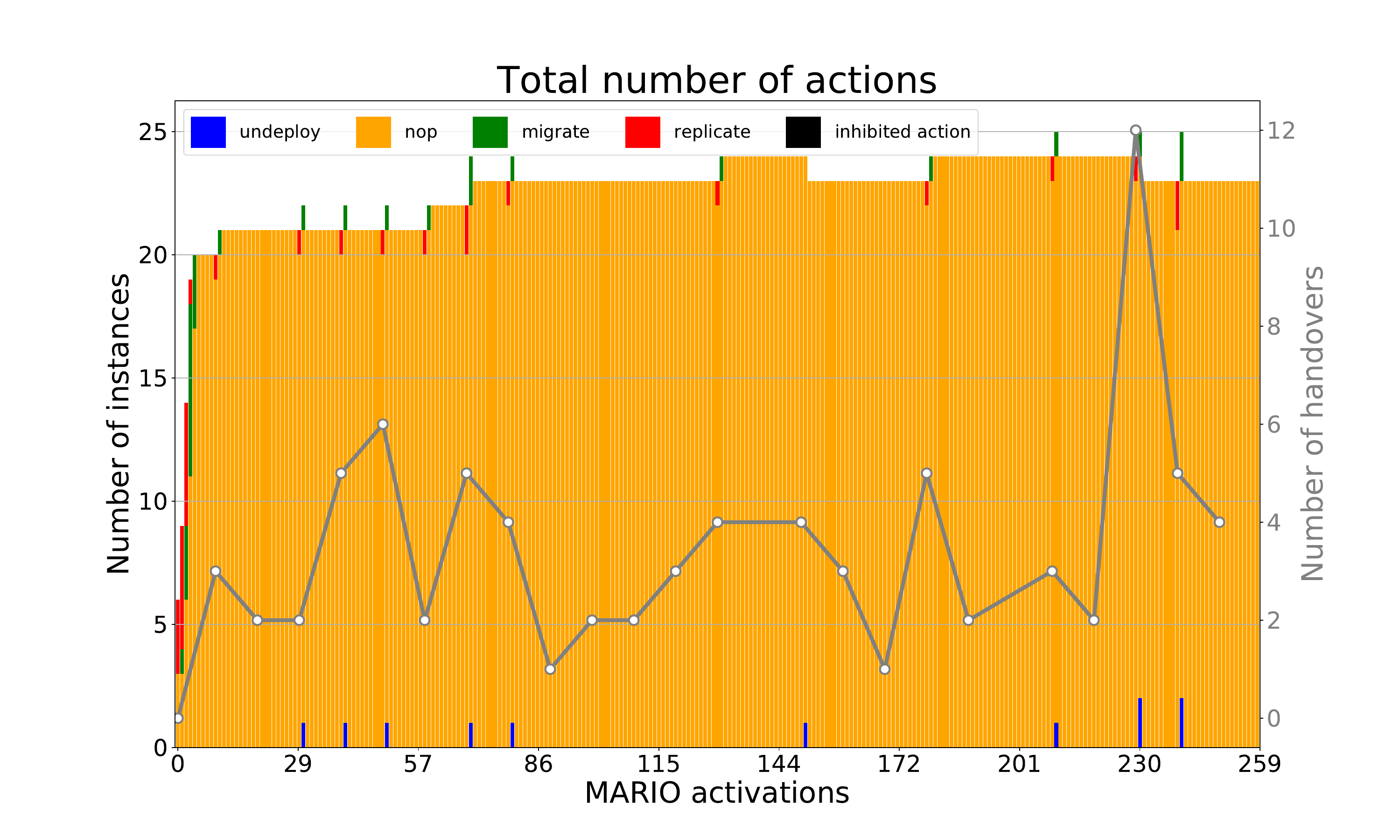}
		\caption{All applications}
		\label{fig:ap2a}
	\end{subfigure}
	\caption{Experiments with \policytwo.}
	\label{fig:actionsP2}
\end{figure}

\paragraph{\policytwo in Action --} Figure~\ref{fig:actionsP2} shows the result when policy \policytwo is used across all applications, i.e. with the goal of migrating and replicating instances in a latency-aware manner. Remember that \ws were defined with a relaxed latency constraint which does not trigger them to move away from the central Cloud (as per the initial setup), even though they are experiencing overloading. On the contrary, the \ls set of applications has a tighter latency constraint, which forces them to move as close as possible to edge nodes as per \policytwo.
%
Indeed, as per Figure~\ref{fig:ap2b}, the three application instances of type \ws are never replicated -- if needed -- nor migrated only incurring in {\tt\small nop} operations. This brings to zero the number inhibited operations, as less resources are consumed in resource-constrained node at the edge of the network. As expected, Figure~\ref{fig:ap2c} shows instead how \ls application instances increase up to 21 throughout the simulation in order to get closer to their users and satisfy their latency constraints. 

Again, as illustrated in Figure~\ref{fig:ap2a}, only 1 or 2 management cycles suffice to reach a temporarily steady allocation in response to user movements. Differently from the experiments with \policyone, in these settings, all application instances achieve their suitable allocation at every step of the simulations as no inhibited operation is observed.

\begin{figure}
	\begin{subfigure}[b]{0.9\textwidth}
		\centering
		\includegraphics[width=0.8\textwidth]{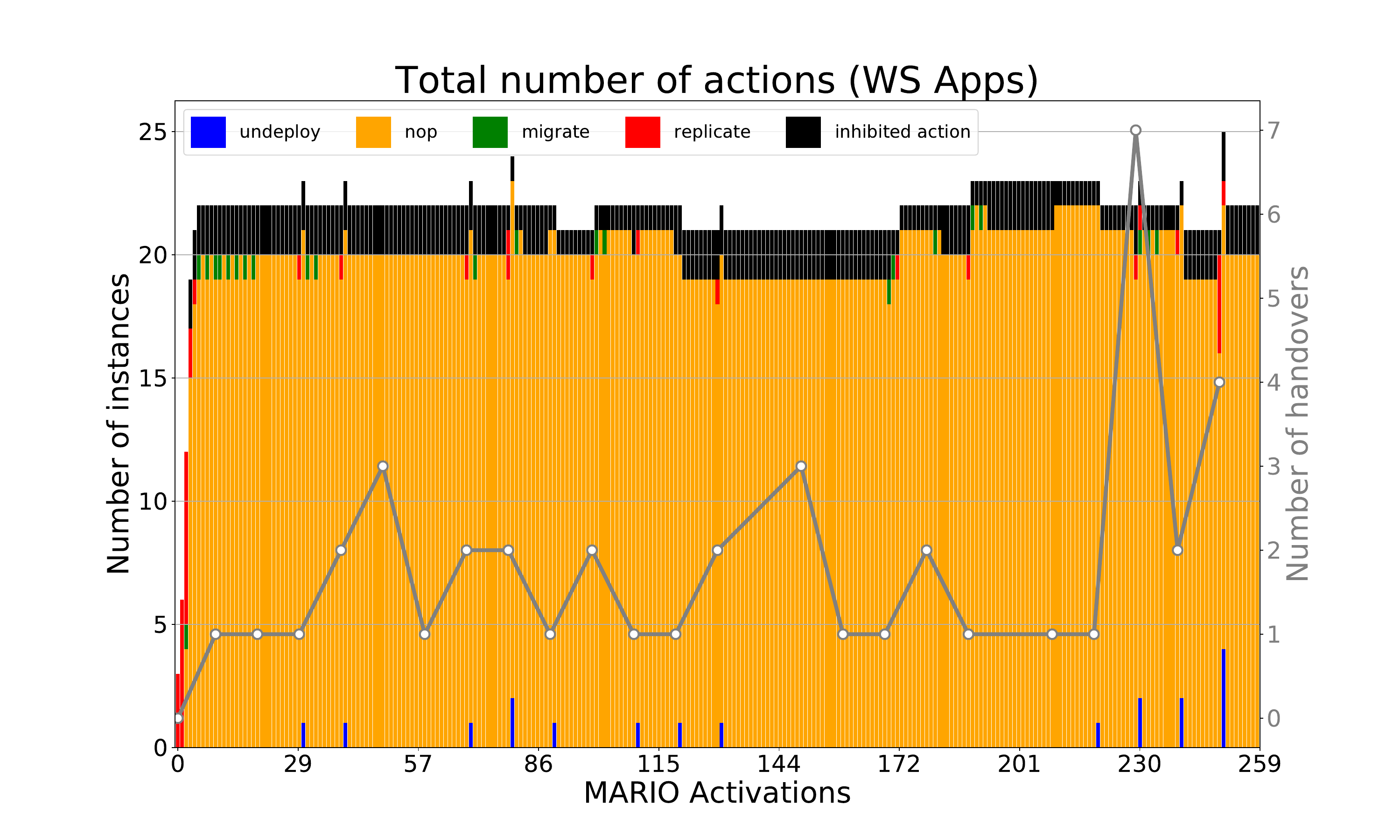}
		\caption{\ws applications}
		\label{fig:ap12b}
	\end{subfigure}
	\hfill
	\begin{subfigure}[b]{0.9\textwidth}
		\centering
		\includegraphics[width=0.8\textwidth]{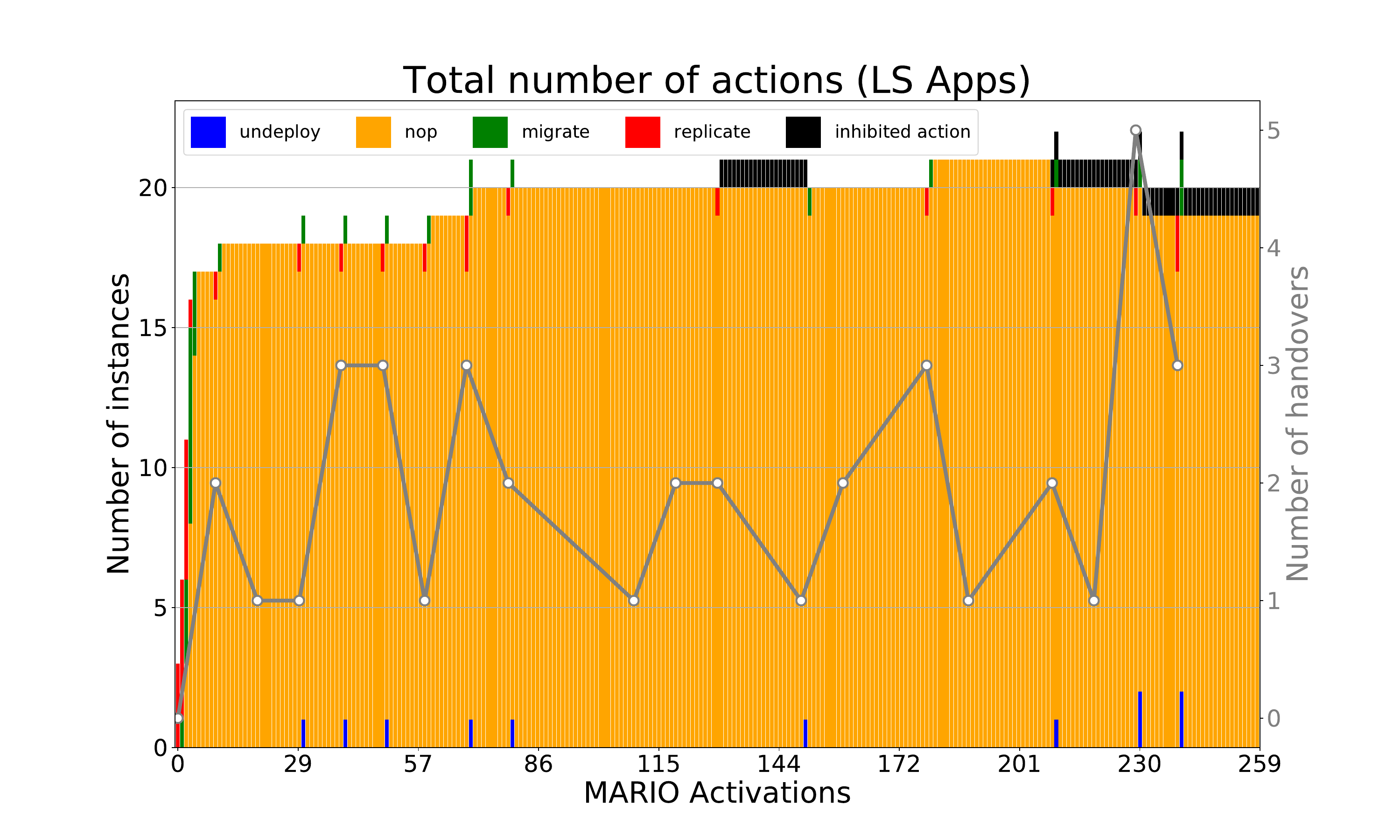}
		\caption{\ls applications}
		\label{fig:ap12c}
	\end{subfigure}
	\hfill
	\centering
	\begin{subfigure}[b]{0.9\textwidth}
		\centering
		\includegraphics[width=0.8\textwidth]{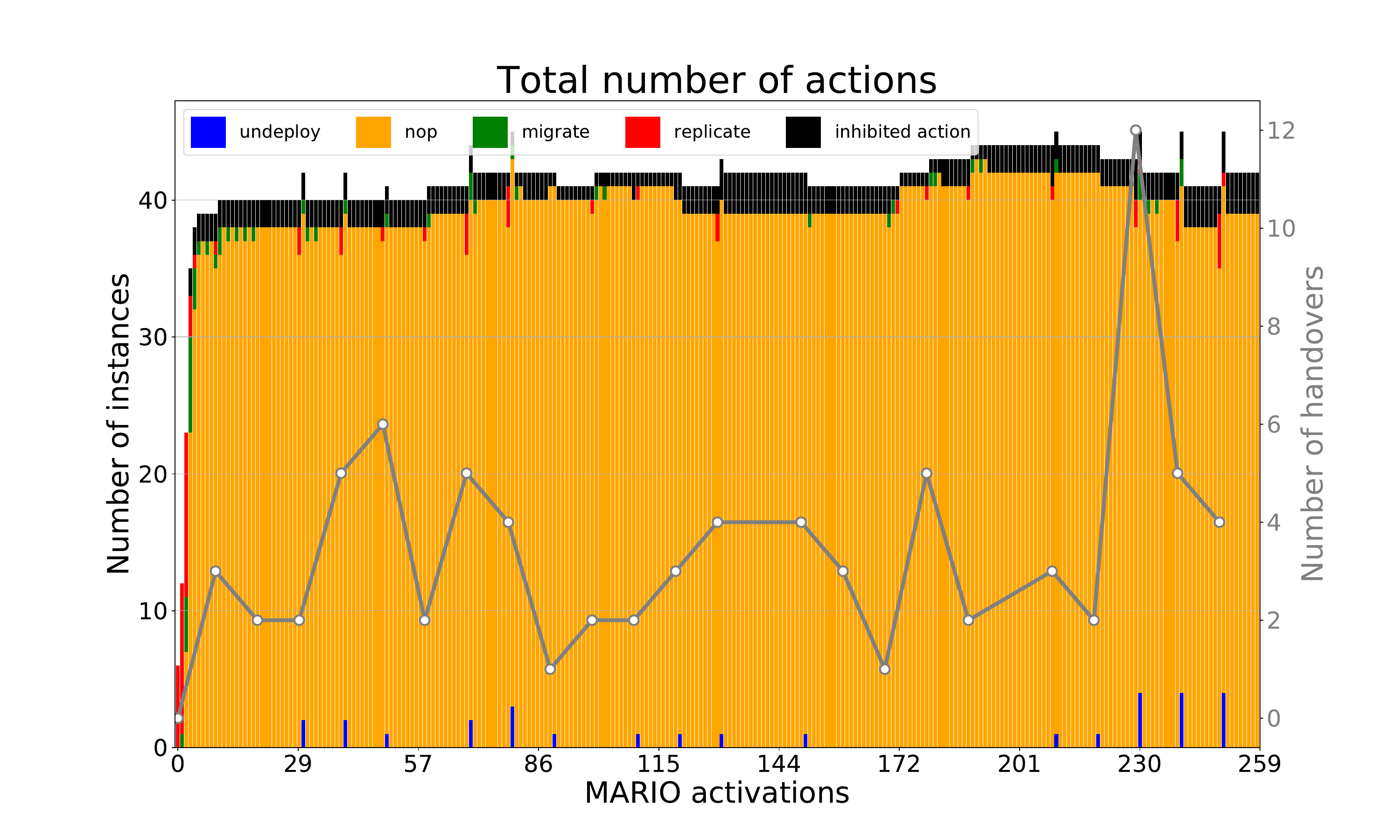}
		\caption{All applications}
		\label{fig:ap12a}
	\end{subfigure}
	\caption{Experiments with \policythree.}
	\label{fig:actionsP12}
\end{figure}

\paragraph{\policythree in Action --} Figure~\ref{fig:actionsP12} shows the results for the third experiment, where \policythree is enforced across all applications. It is worth recalling that this policy is both workload-aware and latency-aware. Consequently, the applications from both set should try be replicated and migrated closer to the users. As shown in Fig.~\ref{fig:ap12b} and Fig.~\ref{fig:ap12c}, both types of applications scale up to 20 instances and more as per \policythree. 
Naturally, as both types of applications are trying to scale, the number of inhibited actions increases in the considered scenario. It is worth noting, however, that as in the experiments with \policyone only (Fig. \ref{fig:actionsP1}), 
operations on \ws applications are inhibited more often than operations on \ls applications
(Fig. \ref{fig:ap12a}). This is due to the fact that trying to replicate on edge nodes is not always possible due to their limited resources. On the other hand, once an application of type \ls reaches a node from which it can guarantee the required latency constraints, it will remain there all through until its users move, thus triggering either a {\tt\small migrate} or an {\tt\small undeploy} operation (Fig. \ref{fig:ap12b}).


As illustrated by Fig.~\ref{fig:ap12a}, despite more application instances need to adapt their allocation in response to user movements, the management cycles needed to react to changes is still reasonably small, being bound by 2 on average and by $6$ in the worst-case (which only happens once in the simulation). 

\paragraph{\policyfour in Action --} The last experiment corresponds to the use of \policyfour, which extends \policythree by avoiding that a recently inhibited management operations is tried again. For the sake of the experiments, we forbid retrying operations that were rejected in the previous management cycle (i.e. we keep in the knowledge base of each management agent only the {\tt\small inhibited/3} facts related to the last policy evaluation) and in the previous $10$ management cycles. 

\begin{figure}
	\centering
	\begin{subfigure}[b]{0.9\textwidth}
		\centering
		\includegraphics[width=0.8\textwidth]{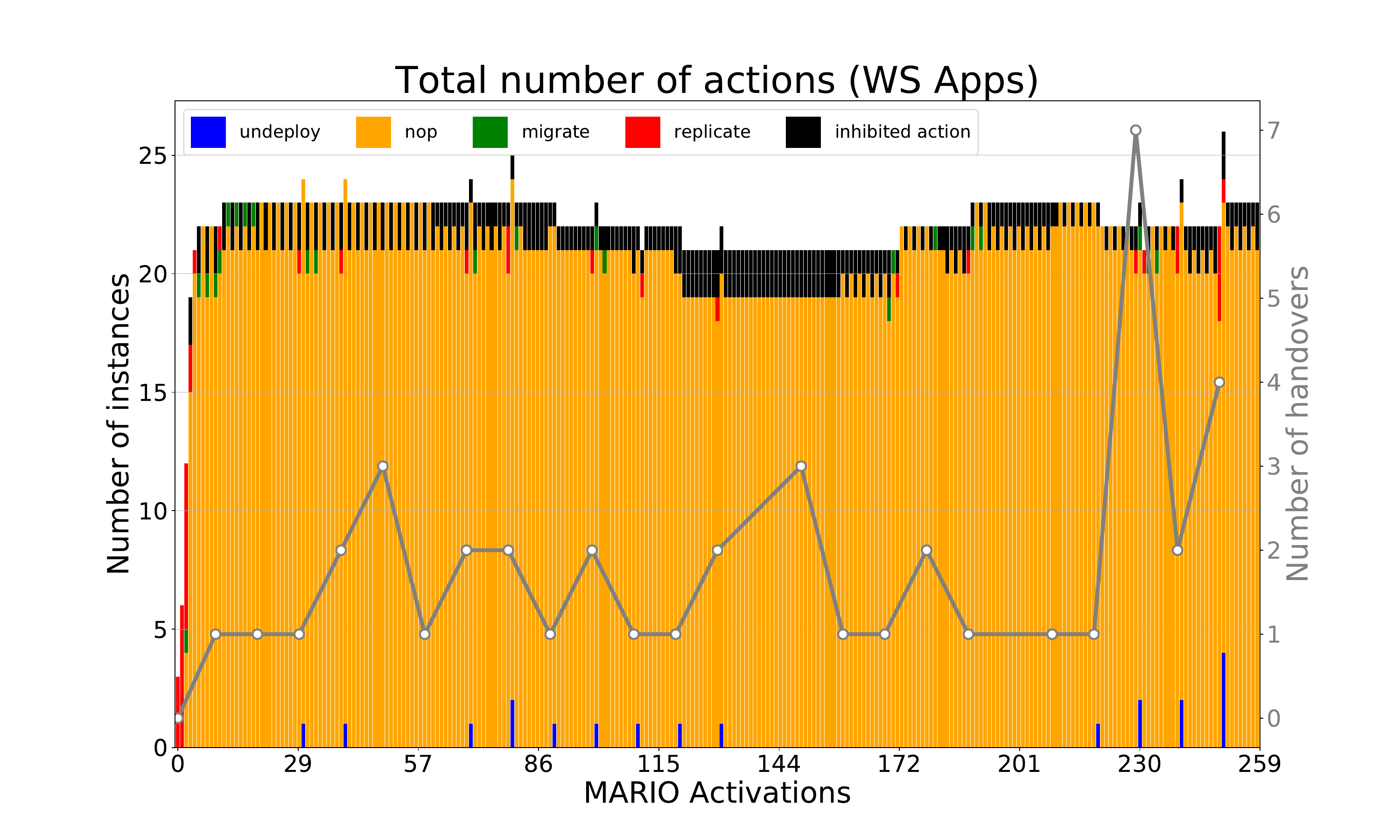}
		\caption{\ws applications}
		\label{fig:ap12Mb}
	\end{subfigure}
	\hfill
	\begin{subfigure}[b]{0.9\textwidth}
		\centering
		\includegraphics[width=0.8\textwidth]{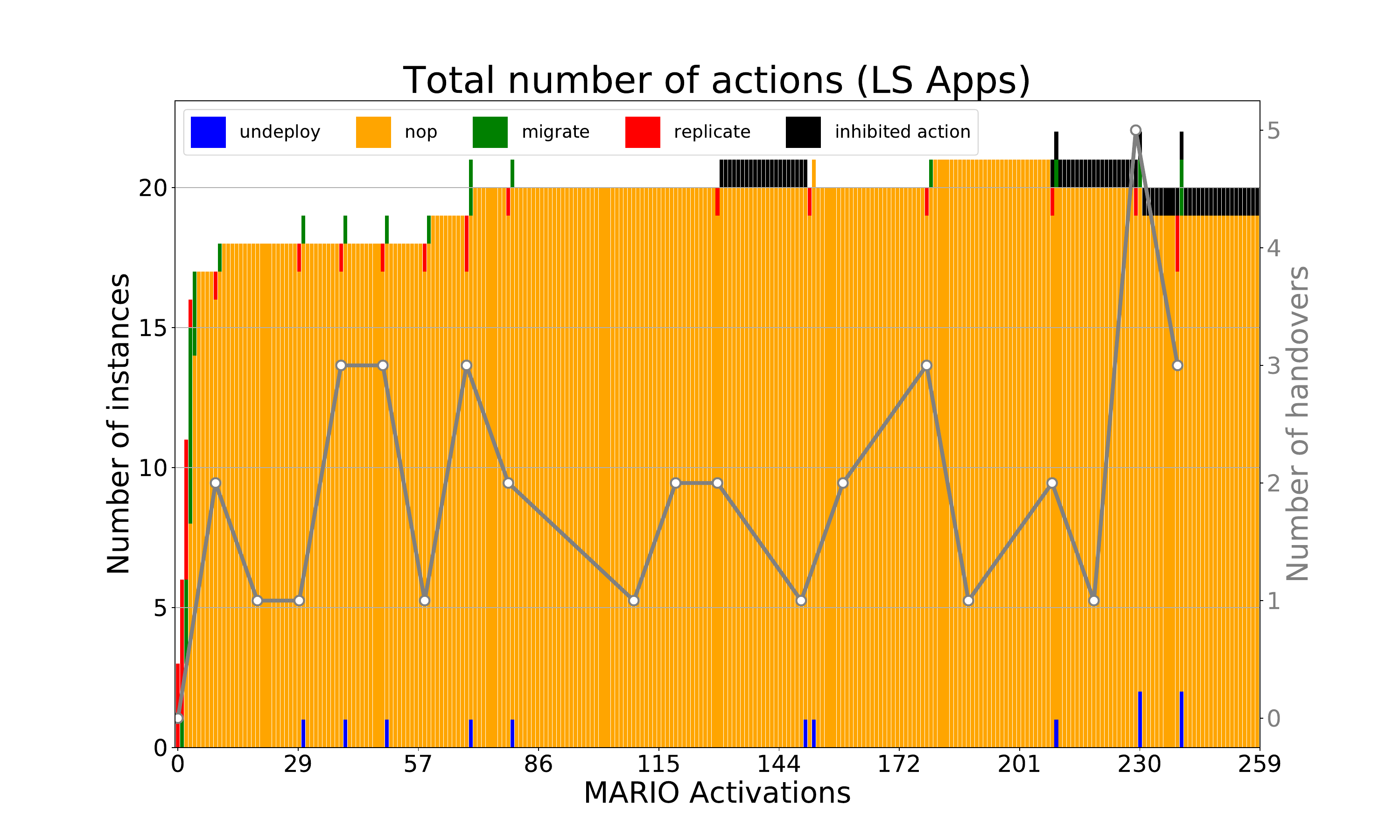}
		\caption{\ls applications}
		\label{fig:ap12Mc}
	\end{subfigure}
	\hfill
	\begin{subfigure}[b]{0.9\textwidth}
		\centering
		\includegraphics[width=0.8\textwidth]{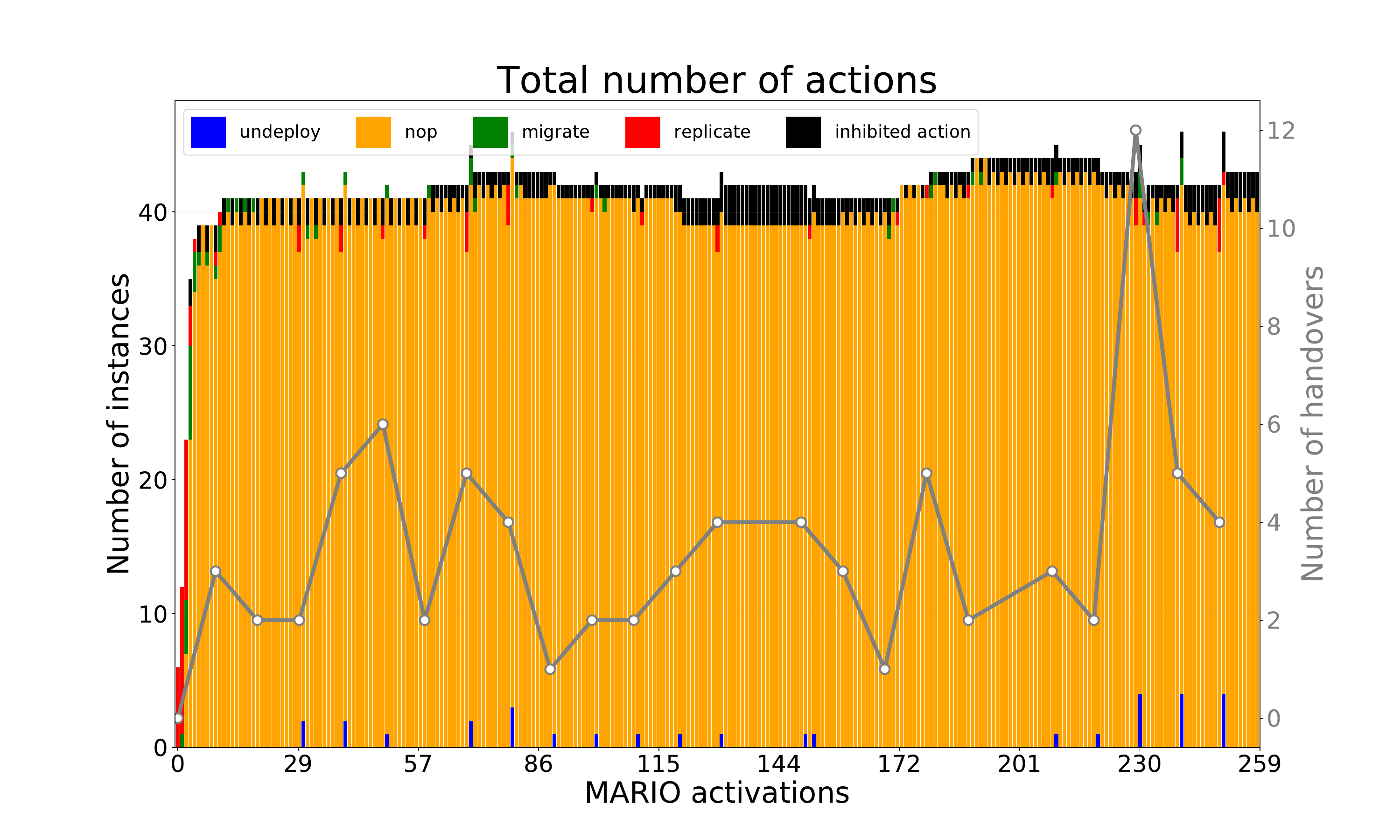}
		\caption{All applications}
		\label{fig:ap12Ma}
	\end{subfigure}
	\caption{Experiments with \policyfour (memory size 1).}
	\label{fig:actionsP12M}
\end{figure}

Even with a small memory of inhibited actions, the general behaviour of \policyfour is very similar to the one of \policythree but it shows a $23\%$ reduction of the number of inhibited operations across all applications (Figure~\ref{fig:actionsP12M}). Indeed, the behaviour of \ws applications plot in Fig.~\ref{fig:ap12Mb} shows that inhibited agents can almost always find an alternative operation in the next management cycle (differently from \policythree, Fig. \ref{fig:ap12b}).
Factually, we can observe that there are fewer consecutive \textit{inhibited} actions throughout the simulation. 
On the other hand, the behaviour of \ls applications is practically unchanged as shown in Fig. \ref{fig:ap12Mc}.
As in all the previous experiments, the management operations needed to react to changes the applications are bound to a small constant, viz. $2$ on average (Figure~\ref{fig:ap12Ma}).

\begin{figure}[]
	\centering
	\begin{subfigure}[b]{0.9\textwidth}
		\centering
		\includegraphics[width=0.8\textwidth]{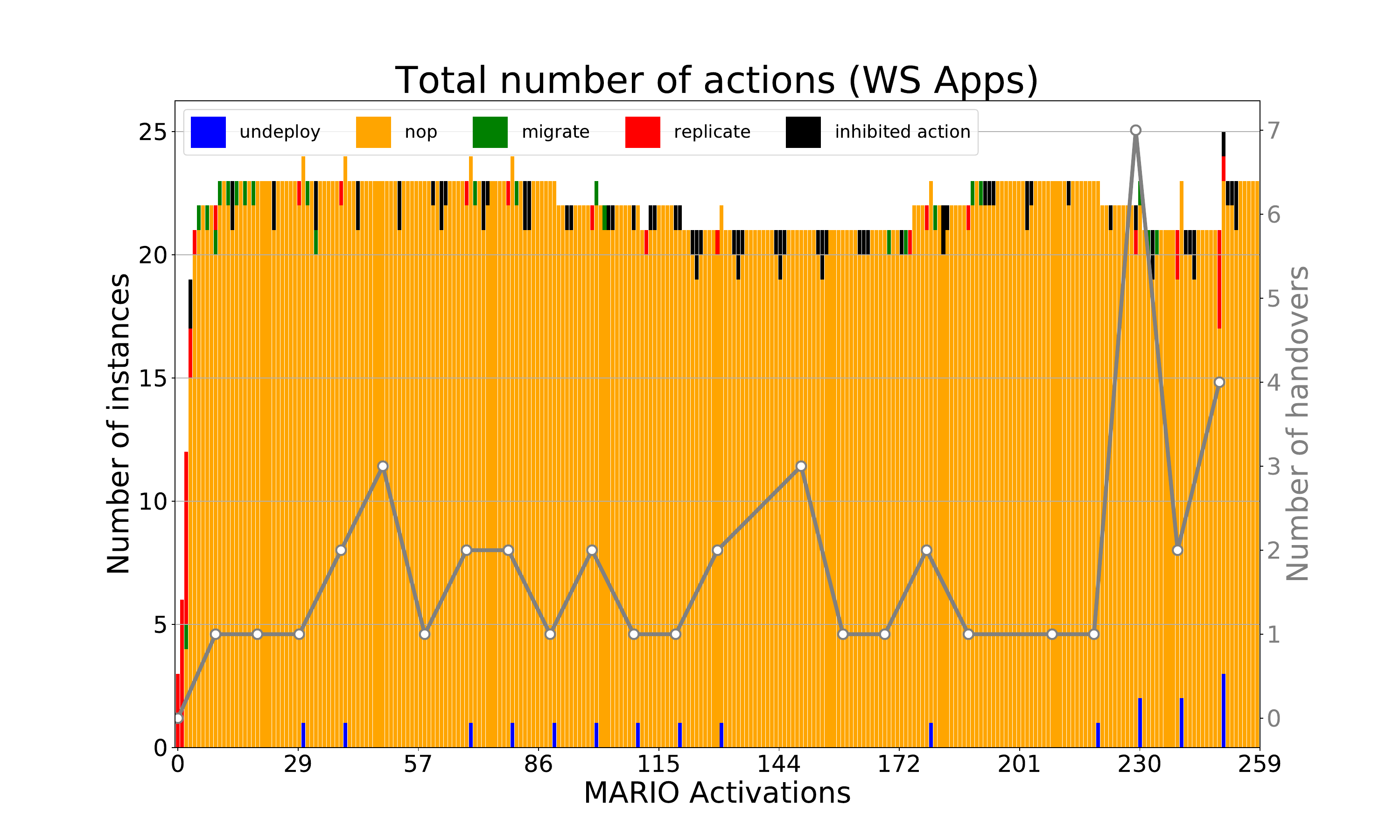}
		\caption{\ws applications}
		\label{fig:ap12Mbw10}
	\end{subfigure}
	\hfill
	\begin{subfigure}[b]{0.9\textwidth}
		\centering
		\includegraphics[width=0.8\textwidth]{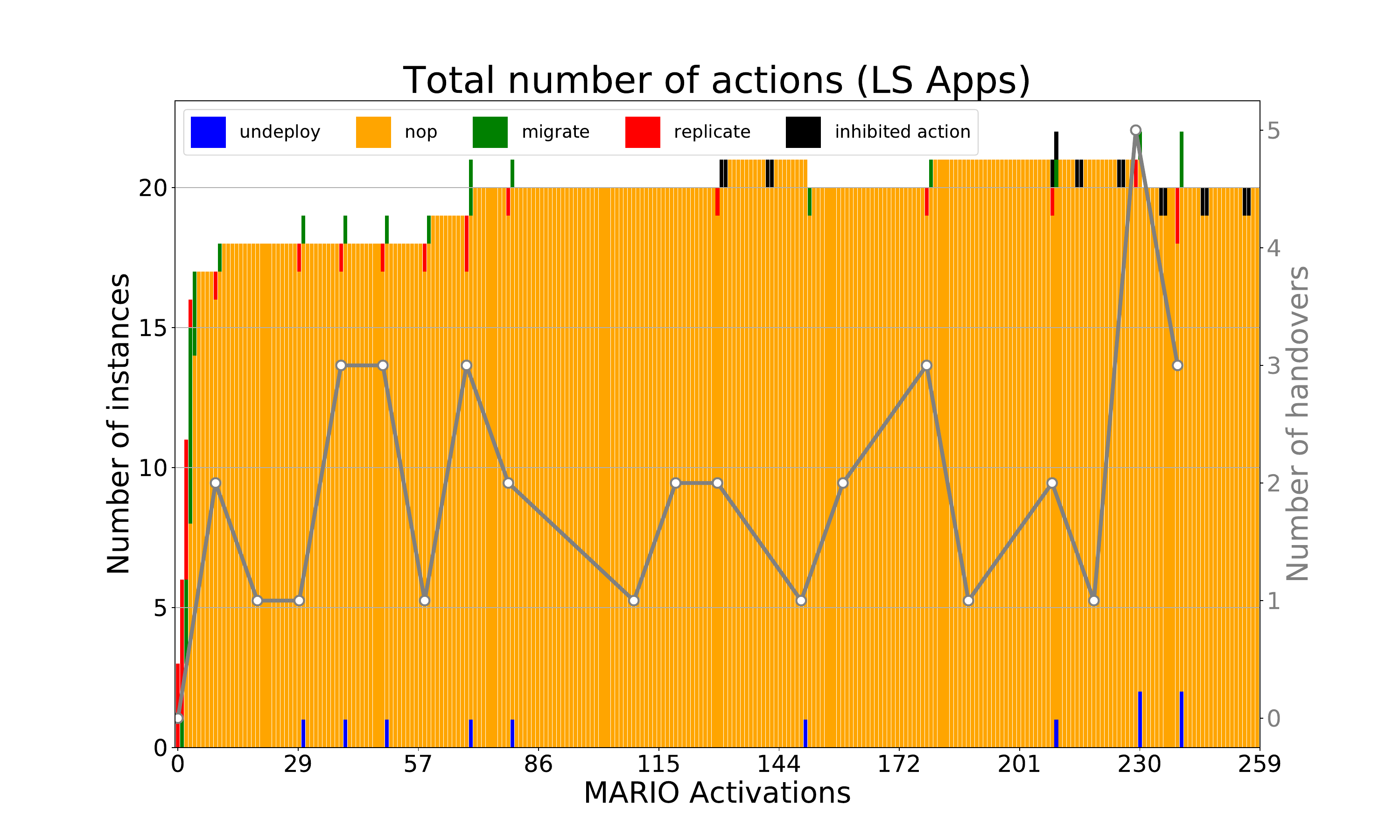}
		\caption{\ls applications}
		\label{fig:ap12Mcw10}
	\end{subfigure}
	\hfill
	\begin{subfigure}[b]{0.9\textwidth}
		\centering
		\includegraphics[width=0.8\textwidth]{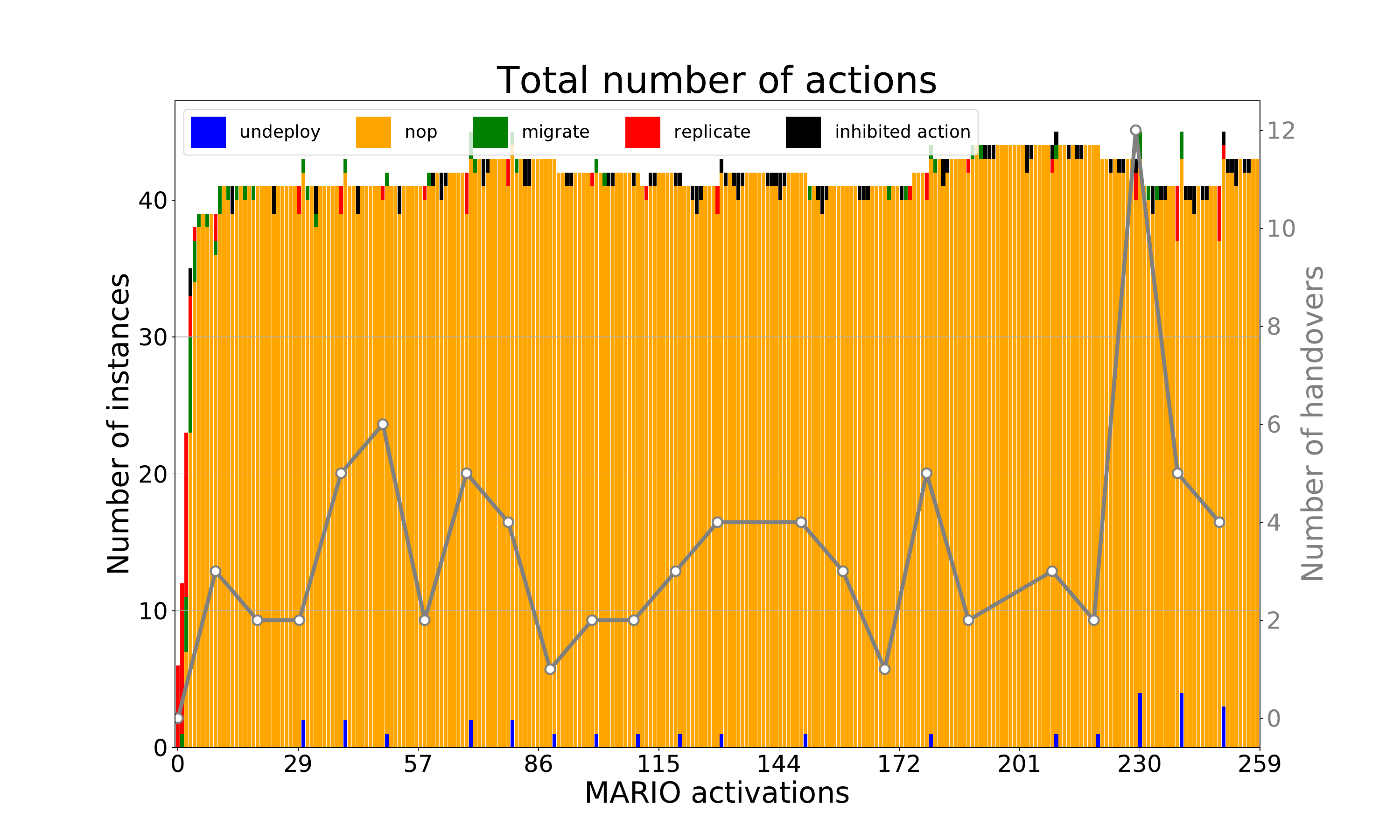}
		\caption{All applications}
		\label{fig:ap12Maw10}
	\end{subfigure}
	\caption{ Experiments with \policyfour (memory size 10). }       
	\label{fig:actionsP12Mw10}
\end{figure}

As expected, increasing the memory size to the last $10$ management cycles further improves these results (Fig. \ref{fig:actionsP12Mw10}). Indeed, \ws applications are now capable of always finding a suitable alternative to the previously inhibited operation in the consecutive management cycle, thus reducing the period needed to react to changes (Fig. \ref{fig:ap12Mbw10}). Similarly, \ls applications incur in a negligible number of inhibited actions throughout the simulation, viz. 16 throughout the simulation (Fig. \ref{fig:ap12Mcw10}).
Such improved emerging behaviour of the system is also shown by Fig. \ref{fig:ap12Maw10}, aggregating on both \ws and \ls applications. Overall inhibited actions are reduced by more than 80\% in this last scenario, with respect to \textit{Policy 3}.

\begin{figure}
	\centering
	\begin{subfigure}[b]{0.45\textwidth}
		\centering
		\includegraphics[width=\textwidth]{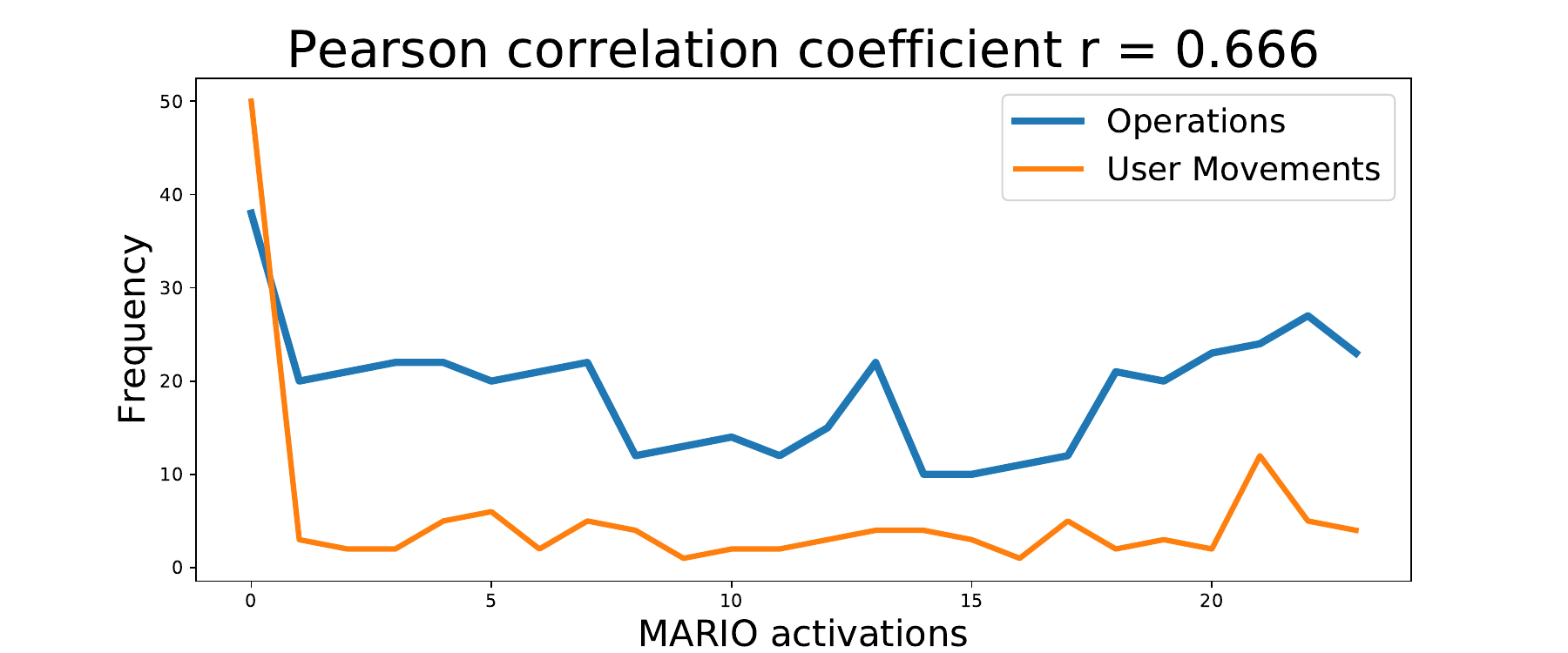}
		\caption{All applications}
		\label{fig:p1a}
	\end{subfigure}
	\\
	\begin{subfigure}[b]{0.45\textwidth}
		\centering
		\includegraphics[width=\textwidth]{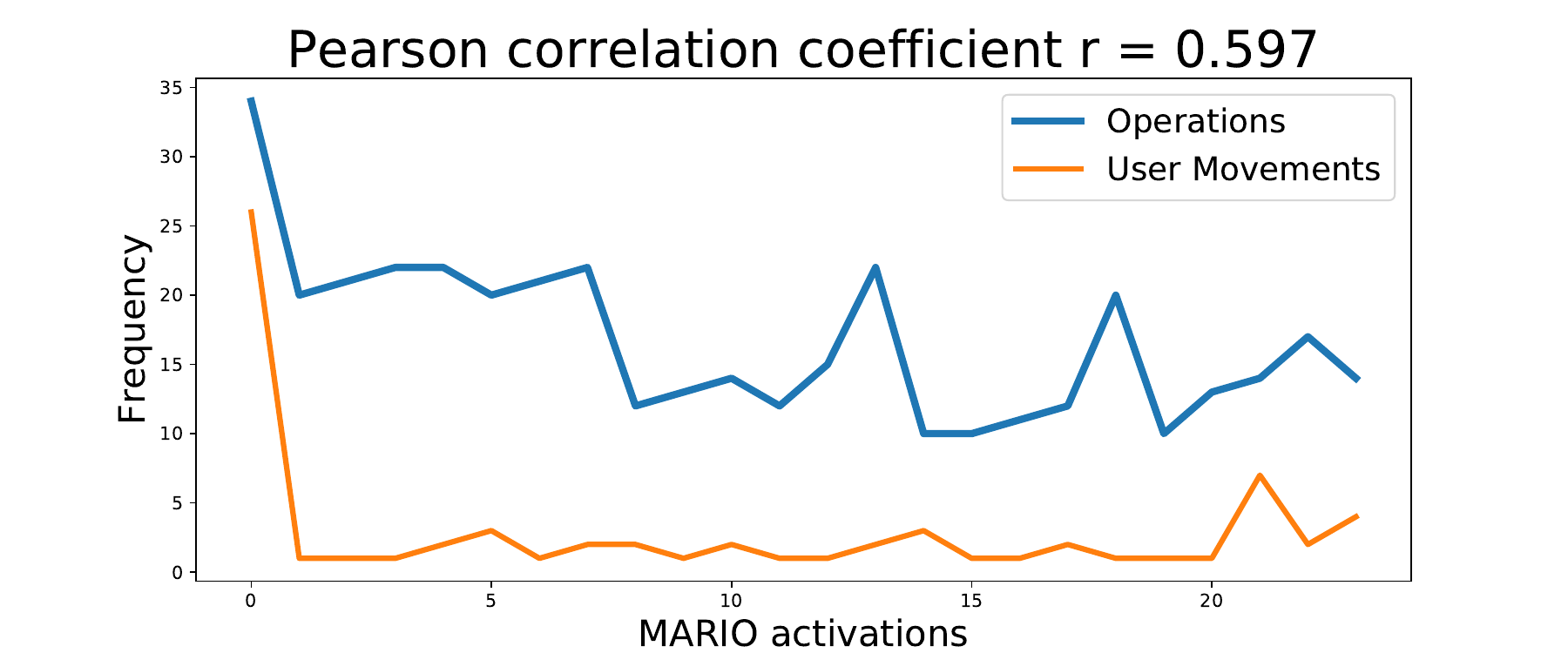}
		\caption{\ws applications}
		\label{fig:p1b}
	\end{subfigure}
	\hfill
	\begin{subfigure}[b]{0.45\textwidth}
		\centering
		\includegraphics[width=\textwidth]{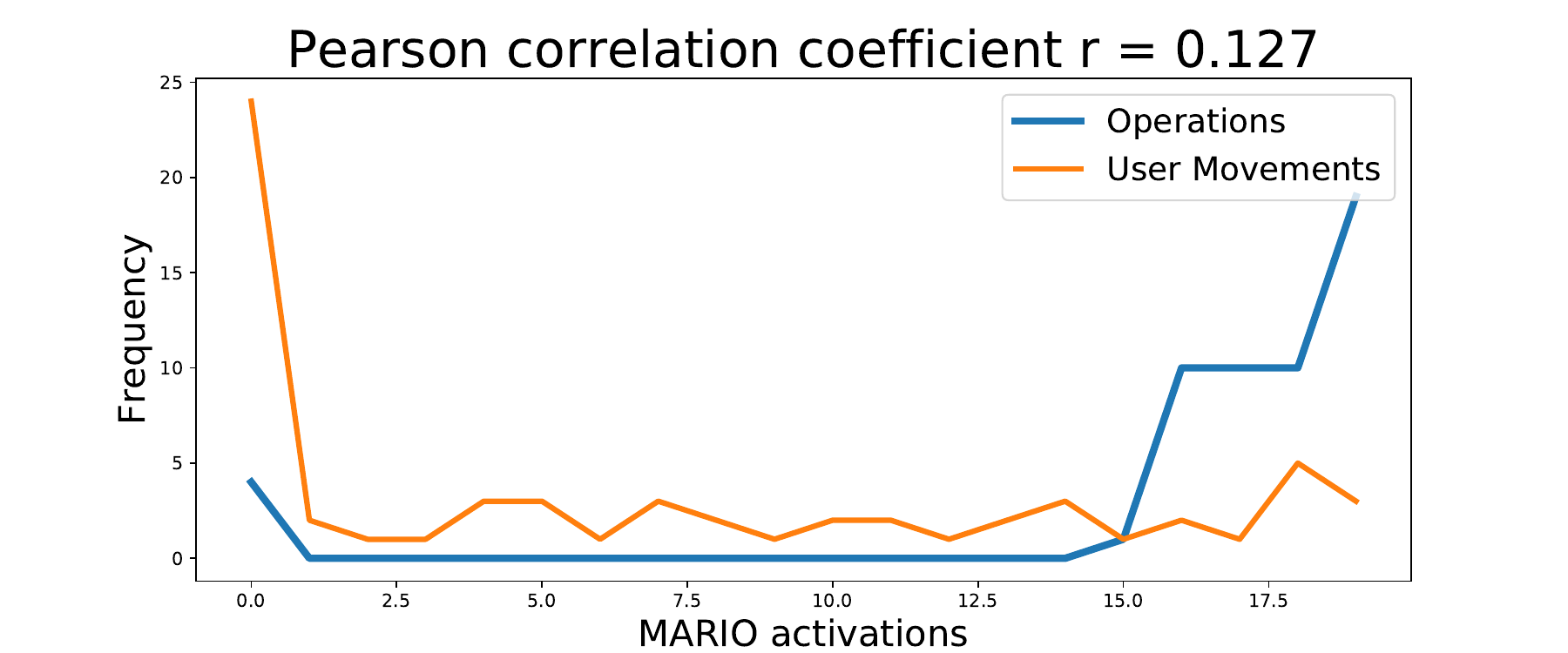}
		\caption{\ls applications}
		\label{fig:p1c}
	\end{subfigure}
	\caption{Pearson correlation coefficient for \policyone.}
	\label{fig:policy1}
\end{figure}

\begin{figure}
	\centering
	\begin{subfigure}[b]{0.45\textwidth}
		\centering
		\includegraphics[width=\textwidth]{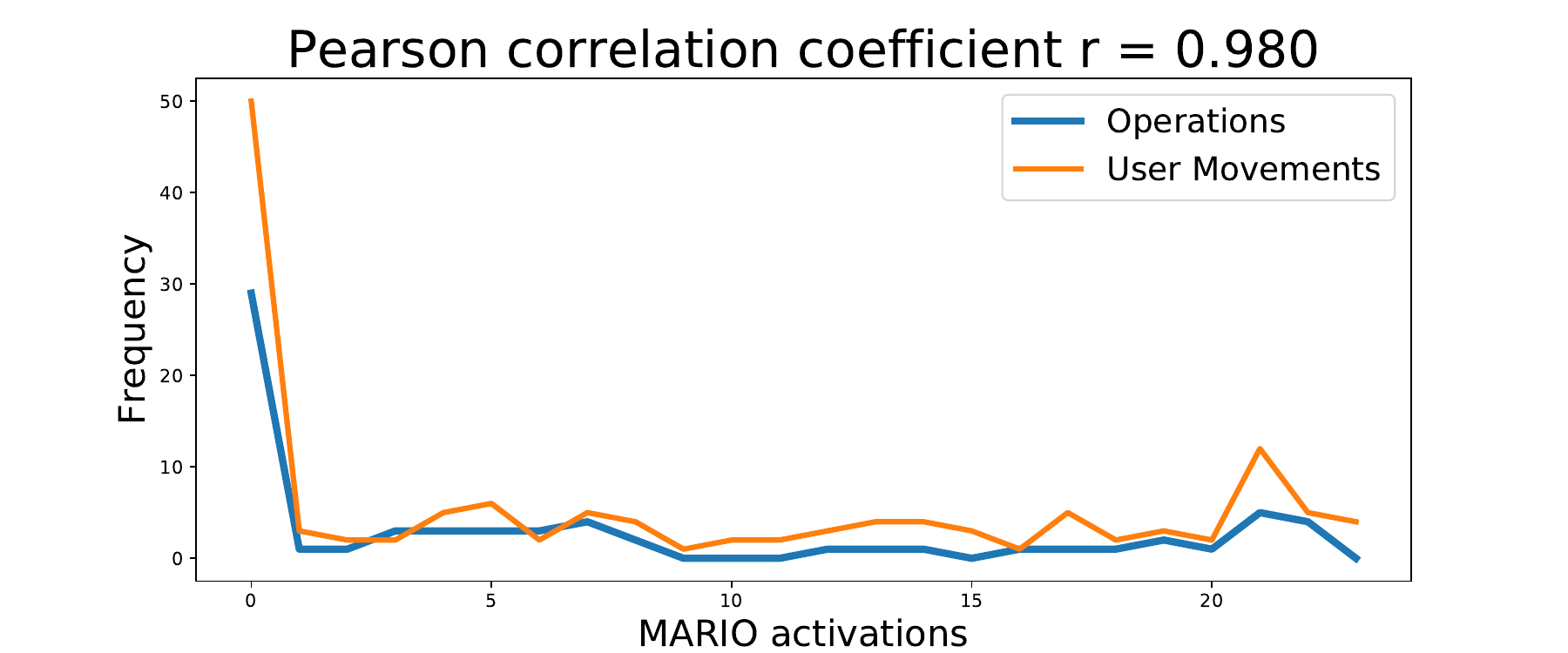}
		\caption{All applications}
		\label{fig:p2a}
	\end{subfigure}
	\\
	\begin{subfigure}[b]{0.45\textwidth}
		\centering
		\includegraphics[width=\textwidth]{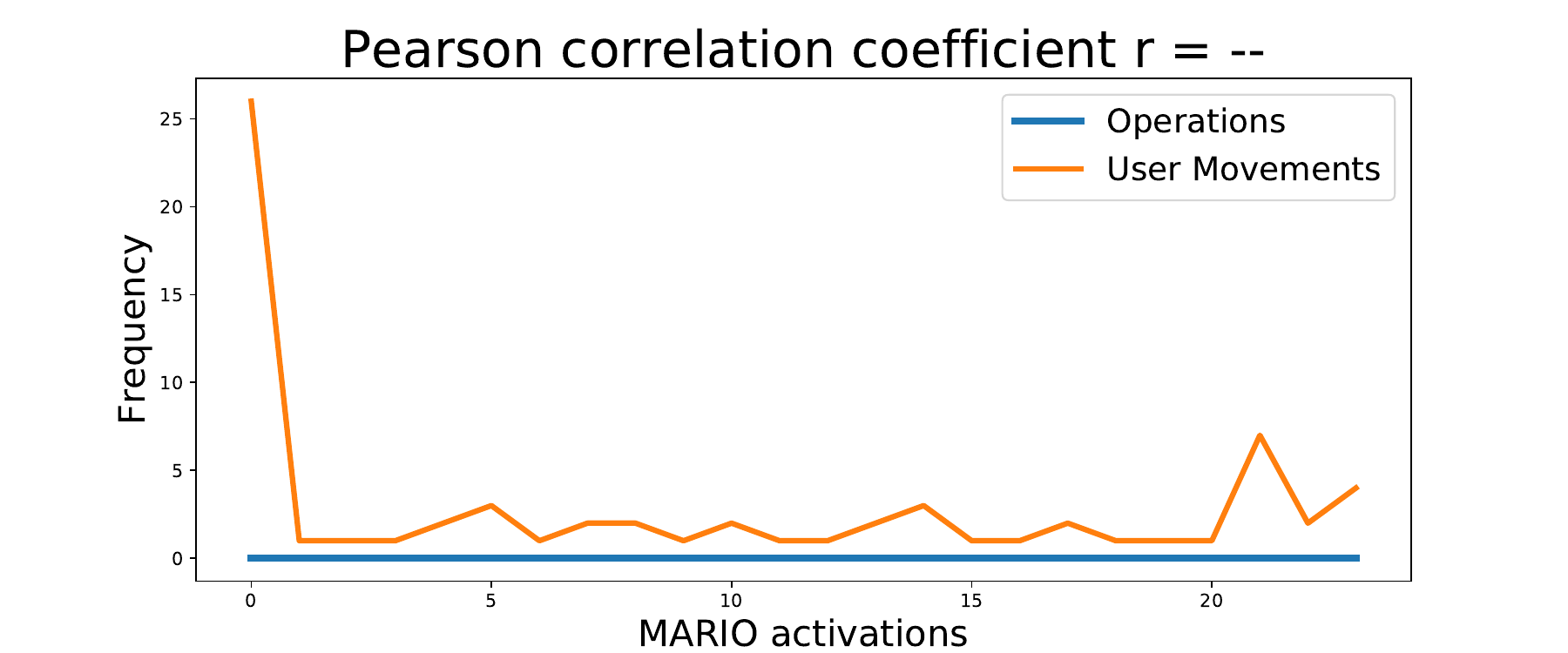}
		\caption{\ws applications}
		\label{fig:p2b}
	\end{subfigure}
	\hfill
	\begin{subfigure}[b]{0.45\textwidth}
		\centering
		\includegraphics[width=\textwidth]{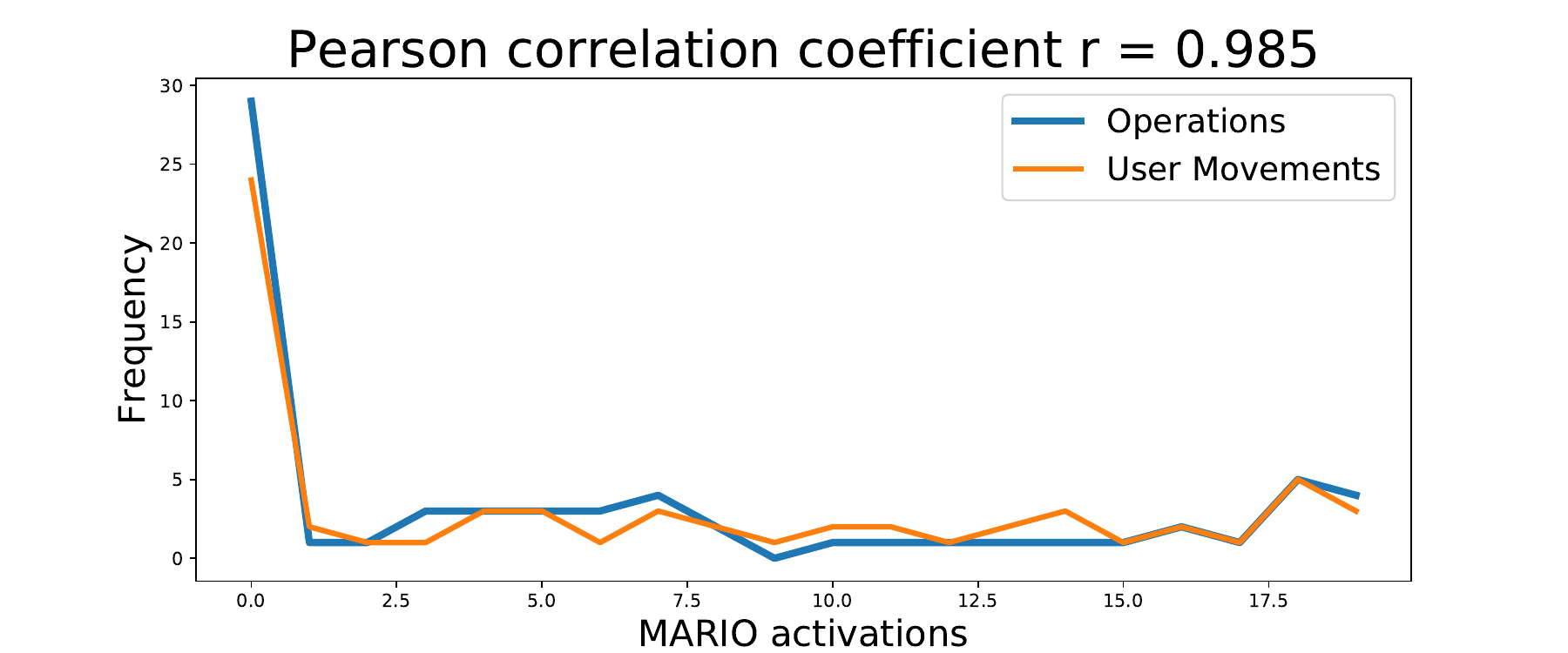}
		\caption{\ls applications}
		\label{fig:p2c}
	\end{subfigure}
	\caption{Pearson correlation coefficient for \policytwo.}
	\label{fig:policy2}
\end{figure}

\begin{figure}
	\centering
	\begin{subfigure}[b]{0.45\textwidth}
		\centering
		\includegraphics[width=\textwidth]{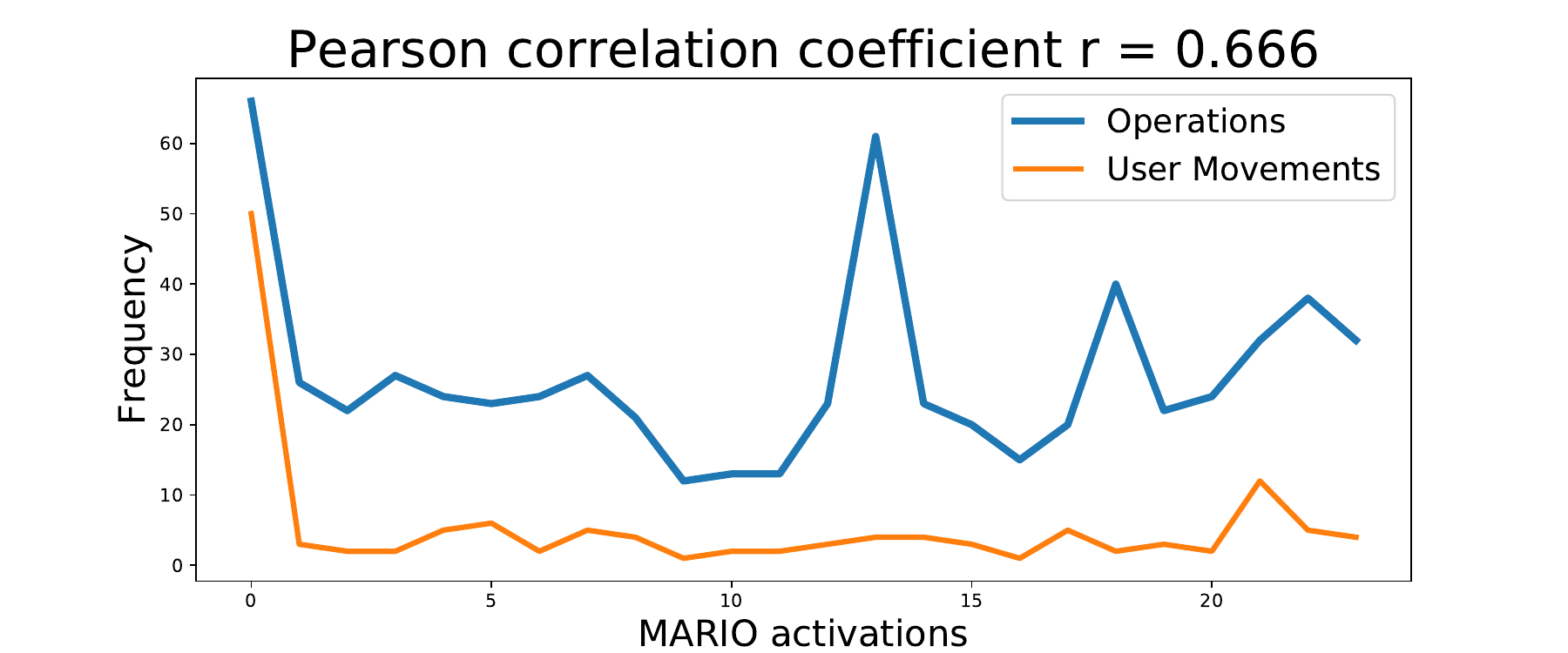}
		\caption{All applications}
		\label{fig:p12a}
	\end{subfigure}
	\\
	\begin{subfigure}[b]{0.45\textwidth}
		\centering
		\includegraphics[width=\textwidth]{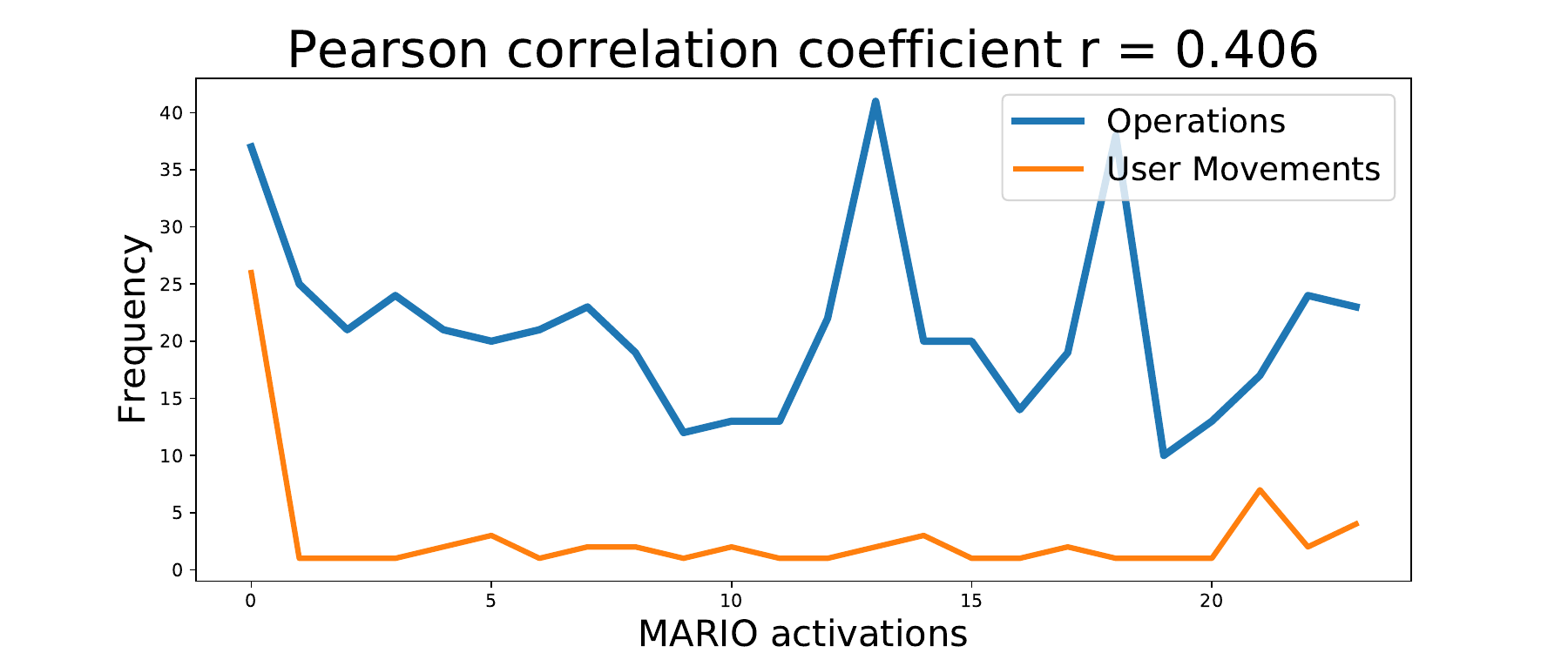}
		\caption{\ws applications}
		\label{fig:p12b}
	\end{subfigure}
	\hfill
	\begin{subfigure}[b]{0.45\textwidth}
		\centering
		\includegraphics[width=\textwidth]{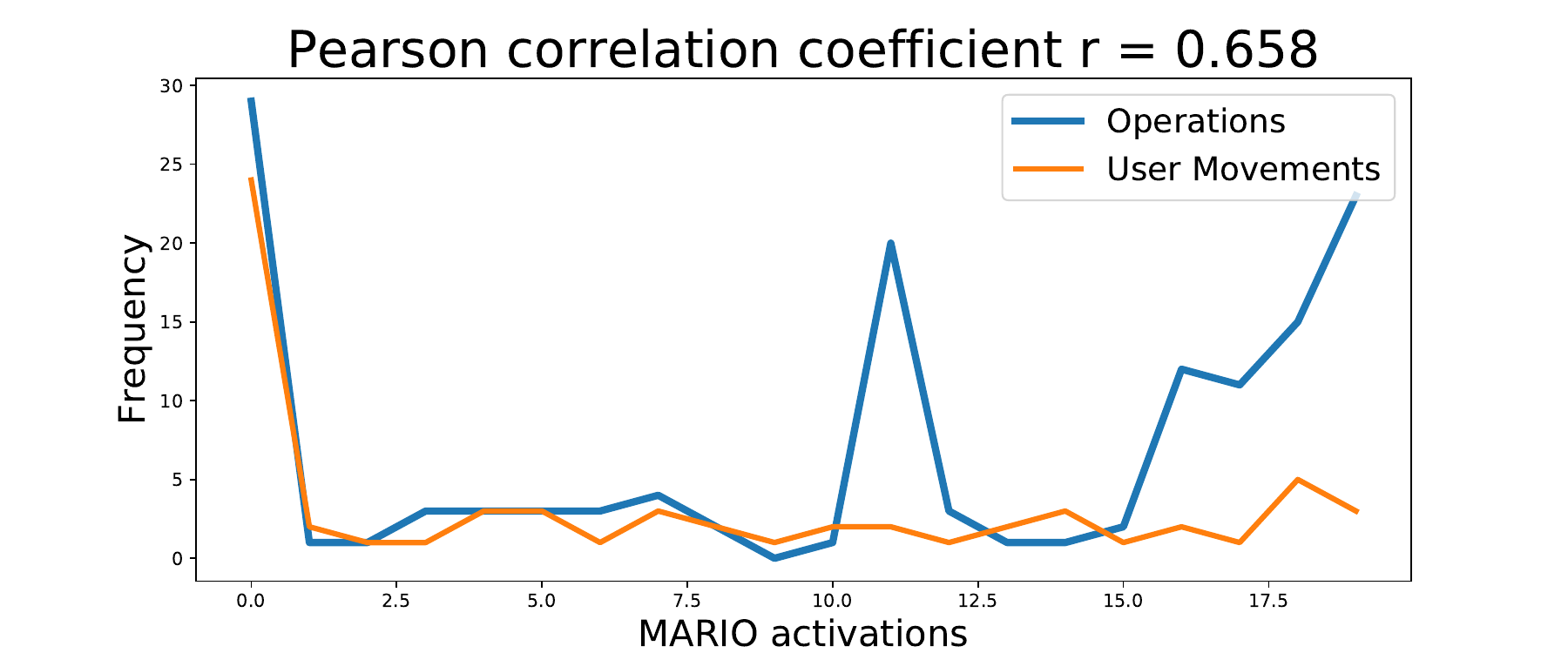}
		\caption{\ls applications}
		\label{fig:p12c}
	\end{subfigure}
	\caption{Pearson coefficient for \policythree}
	\label{fig:policy12}
\end{figure}

\paragraph{Correlating Management Operations and User Movements --} Additionally, we have analysed how the number of user movements influences the number of {\tt\small migrate}, {\tt\small replicate} or {\tt\small undeploy} operations. Expectedly, the higher the number of user movements, the higher the number of operations that change the allocation of the services. We have measured this influence by calculating the Pearson's correlation coefficient ($r$) between the series that counts the number of user movement for each period and the series that counts the number of {\tt\small migrate}, {\tt\small replicate} or {\tt\small undeploy} operations in those same periods. These correlation evaluations have been performed independently for each set of applications (\ws and \ls) and for the aggregation of both sets. Figures~\ref{fig:policy1}--\ref{fig:policy12} plot the two series for each case, and the resulting Pearson's correlation coefficient.

Generally speaking, there is a strong correlation between the movement of the users and the number of management operations performed in the system. However, such correlation is only present when management policies try to allocate applications closer to the users. This is the case, for example, of \ws applications managed via \policyone ($r=0.597$, Fig. \ref{fig:policy1}) and of \ls applications managed via \policytwo ($r=0.985$, Fig. \ref{fig:policy2}). On the contrary, the correlation is very low for the application set that satisfied the constraints by just being allocated further from the users, for example in the Cloud, requiring almost no operation. This is the case of \ls applications managed via \policyone and of \ws applications managed via \policytwo. Notice that correlation coefficient is not indicated in Figure~\ref{fig:p2b}. This is because one of the series has always value 0, so coefficient cannot be calculated.

\paragraph{Management Operations per User Movement --} We measured the cost, in terms of number of operations, generated by each user movement. To do that, we have calculated the average number of operations per user movement by dividing the total number of operations of one type by the total number of user movements. We have limited this analysis to the case of the operations that generates changes in the allocation of the services ({\tt\small migrate}, {\tt\small replicate} or {\tt\small undeploy} operations). Table~\ref{table:avgactions1} shows the values separated for policies, operations and applications types.  

\noindent
It is worth noting that the number of operations to react to changes and adapt the application deployment is very low on average, i.e. $\leq 1$ for all policies. 
Consequently, as aforementioned, the cost in terms of management operations is indeed very low and permits fast convergence in practice for the considered lifelike scenario.

\begin{table}[h]
	\caption{Average number of actions by movement.}
	\label{table:avgactions1}
	\centering
	\scriptsize
	\begin{tabular}{|l|lll|lll|}
		\hline Action & \multicolumn{3}{c|}{\textbf{Policy 1}} & \multicolumn{3}{c|}{\textbf{Policy 2}}\\\cline{2-7} 
		& All & \ws Apps & \ls Apps & All & \ws Apps & \ls Apps   \\\hline\hline 
		\textbf{replicate} & 0.548  & 0.930  & 0.128  &  0.341 & 0 &0.717 \\\hline 
		\textbf{migrate} &0.097 &  0.186 & 0         & 0.341 & 0 &0.717\\\hline 
		\textbf{undeploy} & 0.256 & 0.488 & 0         &0.134  & 0 & 0.282\\\hline\hline\hline
		
		\hline Action & \multicolumn{3}{c|}{\textbf{Policy 3}} & \multicolumn{3}{c|}{\textbf{Policy 4}}\\\cline{2-7} 
		& All & \ws Apps & \ls Apps & All & \ws Apps & \ls Apps   \\\hline\hline 
		\textbf{replicate} & 0.792  &0.860  &0.717  &  0.829 & 0.906 &0.743 \\\hline 
		\textbf{migrate} & 0.621 &  0.534 &0.717  & 0.609 &0.534 &0.692\\\hline 
		\textbf{undeploy} & 0.353 & 0.418 & 0.282  &0.378  & 0.441 & 0.307\\\hline 
		
	\end{tabular}
\end{table}

\begin{center}
	\begin{table}[t]
		\caption{Experiment performance results.}\label{tab:experimentResults}
		\centering
		
		\begin{tabular}{|p{3.5cm}|cc|cc|}
			\hline
			
			& \multicolumn{2}{c|}{\textbf{Avg. Response Time (ms)}} & 
			\multicolumn{2}{c|}{\textbf{Service usage}} \\[2pt]\cline{2-5}\cline{2-5}
			\textbf{Experiment} & \textbf{WS Apps} & \textbf{LS Apps} &  \textbf{WS Apps} & \textbf{LS Apps}\\[2pt]\hline
			
			\textbf{\policyone} \newline \textit{Workload-aware} & 3.104 & 8.611 &  1.117 & 0.592  \\\hline
			\textbf{\policytwo} \newline \textit{Latency-aware} & 15.757 & 2.628  & 8.750 & 0.302 \\\hline
			\textbf{\policythree} \newline \textit{Workload- and Latency-aware} & 2.727 & 2.716  & 1.165 & 0.318 \\\hline
			\textbf{\policyfour} \newline \textit{Workload- and Latency-aware,  with  Memory} & 2.590 & 2.652  & 1.131 & 0.303 \\\hline 
		\end{tabular}
		
	\end{table}
\end{center}

\paragraph{Policies and Fog System Performance -- }
Each rule is designed to compensate or balance applications in terms of their sensitivity to latency and workload. We have carried out an analysis of the average response time of each type of application and also a calculation of the level of overload due to the ratio of requests that each application can attend. More precisely, service usage is measured as the ratio between the requests received and the total number of requests that can be served by a service.

The average values of response times and service usages are shown in Table~\ref{tab:experimentResults}. 
The first two rows of Table~\ref{tab:experimentResults} highlight how  \policyone gives preference to workload-sensitive applications, while \policytwo gives preference to latency-sensitive apps. 
As for response time, \policyone prioritizes WS apps and its response time is 3.104 ms, while these applications suffer a response time increase of 407\% with  \policytwo.
LS apps that benefited from  \policytwo have an improvement of 227\%. \policythree and \policyfour compensate for both latency and load characteristics. Both types of applications get fairly similar response times. It is woorth noting that a balance between both application types does not suppose the best assumption and therefore the LS apps do not obtain the best result compared to  \policytwo. Furthermore, \policyfour also represents an improvement compared to \policythree.

In relation to the service usage, the values of \policyone and  \policytwo are again appreciated in the two types of applications. For example, the WS Apps with \policytwo is highly overloaded with a value of 8.750 and this affects the response time. On the contrary, \policytwo compensates LS apps by generating more instances than in \policyone and thus usage service level is reduced and consequently response time improves. \policythree and \policyfour balance the service usage of both types of applications compared to the previous two policies.

\subsection{Scalability}
Generally speaking, a solution is considered to be scalable if the complexity of the problem does not increase when its size does (e.g. number of infrastructure nodes, number of services, number of users).


In comparison with centralised solutions, in \mario each application instance only considers locally available monitored data about the neighbour infrastructure and only data about user requests directly targeting it\footnote{ Note that the number of neighbours of each node (5 in our experiments) depends on the degree of the network graph, not on its size. }. This considerably reduces the amount of information that agents have to process to make informed management decisions at runtime. 
While the management-related problems that agents need to solve are still worst-case exp-time (i.e. placement, scaling, migration), solving them over small instances -- with few 10s of nodes -- is always (exhaustively) feasible as compared to solving them in large-scale settings -- with 100s of nodes~\cite{brogi2019meet}. Additionally, it is important to note that the agents evaluate the management rules in parallel due to their distributed and decentralized execution. 

Summing up,  we can say that \mario succeeds in mastering the exp-time complexity of large-scale Fog infrastructures in that:
\begin{itemize}
	\item The time needed by an infrastructure to “act” (by accepting/rejecting bacteria’s requests) is independent of the number of bacteria in the nodes. All bacteria generate their proposed actions in parallel, and each proposal in this stream is accepted/rejected by the infrastructure nodes in negligible time.
	\item The time needed by the infrastructure nodes to “act” (by accepting/rejecting bacteria’s requests) does not change if we add an arbitrary number of nodes to the infrastructure (but for the case of fully connected infrastructures, which are however not feasible in the Fog domain).
\end{itemize}

\section{Related Work}
\label{sec:related}
\noindent
Much literature has focussed on the placement, orchestration and management of application services to physical servers in Cloud datacentres~(e.g. \cite{pietri2016mapping,TomarchioCM20}). However, only few authors employed declarative programming (e.g. \cite{kadioglu2016heterogeneous,yin2009rhizoma}) or decentralised approaches (e.g. \cite{carlini2016self,sathiaseelan2017towards}), and none combined both.
Moreover, managing applications over the Cloud-IoT continuum introduces new peculiar challenges related to infrastructure scale, lower resource availability, churn and heterogeneity, and to the specificity of IoT application needs, which were rarely considered in Cloud-only scenarios~\cite{ferrer2019towards,edgedeployment}.

Many authors tackled the problem of (sub-)optimally placing and managing services along Cloud-IoT infrastructures. However, few approaches are decentralised.  We refer the readers to recent surveys~\cite{brogi2020place,applicationmanagementfogsurvey,VAQUERO201920} for further details on these topics, and we hereby focus on proposals tackling decentralised and/or declarative approaches to realise next-gen orchestration of Fog applications.

Aiming at collaborative distributed application deployments based on contextual data, \cite{COLISTRA201498} proposed a middleware and a consensus protocol (via broadcast or gossip communication) to achieve decentralised management according to a single non-declarative policy targeting load-balancing. Further work on \mario could investigate embedding a consensus protocol inspired from theirs, to decide when management operations involve more nodes.
\cite{greco16} proposed the GRECO framework, which relies upon distributed genetic algorithms to determine application placements which are likely to resist infrastructure churn, in a failure-resilient manner, by also considering backup application replicas. However, despite the algorithm runs on distributed workers, decision-making still relies upon global knowledge on the available infrastructure.

More recently, \cite{guerrero2019lightweight} proposed a decentralised placement and migration strategy to reduce client-service hop distances. Their work aims at keeping services with lower request rates farther from their users, and services with higher request rates closer to their users. Experimental assessment via simulation has shown promising results in terms of latency reduction and bandwidth usage, with respect to a centralised policy. However, a single non-declarative policy -- which can be expressed also in \mario -- is used for managing applications.
Considering horizontal and vertical application scaling operations, inspired from biological systems, \cite{bioinspired2020} proposed a platform for self-managing containers to optimise load-balancing in a fully decentralised manner. Their approach leverages a single non-declarative policy based on a normalised score accounting for the usage of different resources (i.e. CPU, RAM, I/O), thus disclosing access to those information differently from \mario.
Similarly, focussing on application scaling, \cite{ROSSI2020161} extended Kubernetes decentralised orchestration with self-adaptation and network-aware placement capabilities using reinforcement learning. However, decisions are still made by a centralised control entity relying on a global view of the system.

\vspace{-0.23mm}
Declarative approaches have been proposed in the past to improve network management (e.g. \cite{hinrichs2009practical}) or to express and document management policies in centralised systems \cite{herden2010declarative}. Only recently, few works exploited logic programming to incrementally determine eligible application placements in Fog scenarios \cite{fogbrain}, to place VNF chains and steer traffic across them \cite{edgeusher},  and to enable the assessment of the security and trust levels of different placements by means of algebraic structures \cite{secfog2019}.  From a different perspective, \cite{harzenetter2019automated} employed a declarative approach to describe application requirements based on OpenTOSCA and to automatically deploy applications on different platforms. 

Recently, Casadei et al. \cite{casadei19} and Pianini et al. \cite{casadei21} proposed a declarative approach to service coordination based on \textit{aggregate computing} \cite{aggregate}, i.e. \textit{Self-organising Coordination Regions} (SCR). SCR aims at managing and coordinating opportunistic resources via a hybrid centralised/decentralised solution, by relying on a self-organising peer-to-peer architecture to handle churn and mobility. Differently from \mario, in SCR, few Leader nodes make placement decisions, and application code need restructurings as per the aggregate computing paradigm.  

From the point of view of simulating Fog environments, besides the YAFS simulator~\cite{lera2019yafs} -- which we used to assess \mario\ -- many proposals exist for simulating management policies in Cloud-IoT scenarios~\cite{margariti2020modeling}, e.g.  iFogSim~\cite{004}, EdgeCloudSim~\cite{sonmez2018edgecloudsim}, and FogDirSim~\cite{fortipagiarobrogi2020}.
All those simulators, despite potentially supporting decentralised management policies, were mainly used to assess the performance of simple centralised policies (e.g. random, first-fit, best-fit) reasoning on the conditions of the whole system.
Finally, tackling problems other than decentralised application management, bacteria-inspired solutions -- such as \mario -- have shown promising results in handling the task allocation for systems with multiple unmanned aerial vehicles \cite{bacteriauav}, Cloud security \cite{ahsan2020applications}, software-defined network communication \cite{chang2018bacteria} and self-organising 5G communication networks \cite{bacteria5g}.

Summing up, to the best of our knowledge, no previous work targeting Fog application management exploited a bacteria-inspired, fully decentralised and declarative approach, which permits expressing and flexibly enforcing \textit{ad-hoc} context-aware management policies for different (classes of) applications.

\section{Concluding Remarks}\label{sec:conclusions}

\noindent In this article, by substantially extending our preliminary previous work \cite{mario1}, we have presented \mario: a bacteria-inspired, declarative and fully-decentralised solution for managing next-gen applications in opportunistic Fog computing settings, when it is not possible to adopt centralised solutions.

After defining an operational semantics of our management solution based on three simple operations (viz. undeploy, migrate, replicate), inspired from the emerging behaviour of bacteria colonies, we have shown how to use them to declare four different management policies by means of an open-source Prolog prototype of the proposed solution. We have shown how \mario allows to easily and declaratively specify \textit{ad-hoc} management policies for different (classes of) applications that might want to promptly scale at increasing workload or to enforce latency constraints.
Finally, we have assessed the proposed management policies over a lifelike scenario which considered clients mobility as per real traces of taxi drivers in Rome, Italy. Experimental results validate the proposed management policies and show that \mario is capable of promptly reacting to changes (i.e. users movements, consequent workload variations) within a few evaluations of the management policies, viz. $1$ on average.


As application management agents are fully decentralised and run independently from each other, \mario can scale independently from the infrastructure size and avoids single points of failure that are typical of centralised approaches. Factually, the crash of a node hosting instances of a certain application does not prevent its other instances from replicating so to compensate for the unavailable ones. This, along with the bacteria-inspired behaviour, makes \mario self-organising, hence more resilient to user/device mobility and churn in general, which are typical of opportunistic Fog computing settings~\cite{opportunisticfog2019}.

Last but not least, \mario allows users to specify \textit{ad-hoc} policies for different (classes of) applications, as epitomised in our examples targeting latency-aware and workload-aware applications. As showcased in our experiments, such different dedicated policies can help different applications to optimise their own objective metrics. Along with possible extensions to the information available to \mario agents (e.g. security, costs), this permits to flexibly accommodating a range of diverse and ever-changing application needs in Cloud-IoT scenarios.



As future work, we intend to:
\begin{itemize}
	\item extend the set of management operations featured by \mario, inspired from other bacteria behaviour (e.g. spore-formation, evolution), so to define new management operations (e.g. 
	service adaptation depending on target node as in Osmotic computing \cite{villari2019osmosis}),
	\item 
	include other dimensions to the modelling of \mario such as bandwidth constraints, security policies or operational costs, and more complex priority relations among management operations, to drive management decisions,
	\item \color{black}implement and assess \mario within an existing orchestration platform (e.g. Kubernetes) to test its functioning over Fog testbeds.\color{black}
\end{itemize}

 \bibliographystyle{ieeetr}
 \bibliography{biblio}

\begin{thebibliography}{10}

\bibitem{habibi2020fog}
P.~Habibi, M.~Farhoudi, S.~Kazemian, S.~Khorsandi, and A.~Leon-Garcia, ``{Fog
  Computing: A Comprehensive Architectural Survey},'' {\em IEEE Access},
  vol.~8, pp.~69105--69133, 2020.

\bibitem{villari2019osmosis}
M.~Villari, M.~Fazio, S.~Dustdar, O.~Rana, D.~N. Jha, and R.~Ranjan,
  ``{Osmosis: The osmotic computing platform for microelements in the cloud,
  edge, and internet of things},'' {\em Computer}, vol.~52, no.~8, pp.~14--26,
  2019.

\bibitem{pham2020survey}
Q.-V. Pham, F.~Fang, V.~N. Ha, M.~J. Piran, M.~Le, L.~B. Le, W.-J. Hwang, and
  Z.~Ding, ``A survey of multi-access edge computing in 5g and beyond:
  Fundamentals, technology integration, and state-of-the-art,'' {\em IEEE
  Access}, vol.~8, pp.~116974--117017, 2020.

\bibitem{filali2020multi}
A.~Filali, A.~Abouaomar, S.~Cherkaoui, A.~Kobbane, and M.~Guizani,
  ``Multi-access edge computing: A survey,'' {\em IEEE Access}, 2020.

\bibitem{brogi2020place}
A.~Brogi, S.~Forti, C.~Guerrero, and I.~Lera, ``{How to Place Your Apps in the
  Fog: State of the Art and Open Challenges},'' {\em Softw. Pract. Exp.},
  vol.~50, no.~5, pp.~719--740, 2020.

\bibitem{applicationmanagementfogsurvey}
R.~Mahmud, K.~Ramamohanarao, and R.~Buyya, ``{Application Management in Fog
  Computing Environments: A Taxonomy, Review and Future Directions},'' {\em ACM
  Comput. Surv.}, vol.~53, no.~4, 2020.

\bibitem{VAQUERO201920}
L.~M. Vaquero, F.~Cuadrado, Y.~Elkhatib, J.~Bernal-Bernabe, S.~N. Srirama, and
  M.~F. Zhani, ``Research challenges in nextgen service orchestration,'' {\em
  Future Gener. Comput. Syst.}, vol.~90, pp.~20 -- 38, 2019.

\bibitem{Ghobaei2020}
M.~Ghobaei-Arani, A.~Souri, and A.~A. Rahmanian, ``Resource management
  approaches in fog computing: a comprehensive review,'' {\em Journal of Grid
  Computing}, vol.~18, no.~1, pp.~1--42, 2020.

\bibitem{fogbrain}
S.~Forti and A.~Brogi, ``Continuous reasoning for managing next-gen distributed
  applications,'' in {\em ICLP Technical Communications 2020}, vol.~325 of {\em
  {EPTCS}}, pp.~164--177, 2020.

\bibitem{guerrero2018migration}
C.~Guerrero, I.~Lera, and C.~Juiz, ``Migration-aware genetic optimization for
  mapreduce scheduling and replica placement in hadoop,'' {\em Journal of Grid
  Computing}, vol.~16, no.~2, pp.~265--284, 2018.

\bibitem{opportunisticfog2019}
N.~{Fernando}, S.~W. {Loke}, I.~{Avazpour}, F.~{Chen}, A.~B. {Abkenar}, and
  A.~{Ibrahim}, ``{Opportunistic Fog for IoT: Challenges and Opportunities},''
  {\em IEEE Internet Things J.}, vol.~6, no.~5, pp.~8897--8910, 2019.

\bibitem{casadeiopportunistic}
R.~Casadei, G.~Fortino, D.~Pianini, W.~Russo, C.~Savaglio, and M.~Viroli,
  ``{Modelling and simulation of Opportunistic IoT Services with Aggregate
  Computing},'' {\em Future Gener. Comput. Syst.}, vol.~91, pp.~252 -- 262,
  2019.

\bibitem{sarteco2019}
I.~Lera, C.~Guerrero, and C.~Juiz, ``Algoritmo descentralizado para la
  asignaci{\'o}n de servicios en arquitecturas de fog computing basado en un
  proceso expansivo de migraci{\'o}n de instancias,'' in {\em Jornadas
  SARTECO}, 2019.

\bibitem{dazzi2020}
P.~Dazzi and M.~Mordacchini, ``Scalable decentralized indexing and querying of
  multi-streams in the fog,'' {\em Journal of Grid Computing}, vol.~18, no.~3,
  pp.~395--418, 2020.

\bibitem{mario1}
A.~Brogi, S.~Forti, C.~Guerrero, and I.~Lera, ``{Towards Declarative
  Decentralised Application Management in the Fog},'' in {\em {GAUSS}}, 2020.
\newblock {\it In press}.

\bibitem{lera2019yafs}
I.~Lera, C.~Guerrero, and C.~Juiz, ``{YAFS: A simulator for IoT scenarios in
  Fog computing},'' {\em IEEE Access}, vol.~7, pp.~91745--91758, 2019.

\bibitem{rometaxis}
L.~Bracciale, M.~Bonola, P.~Loreti, G.~Bianchi, R.~Amici, and A.~Rabuffi,
  ``{CRAWDAD dataset roma/taxi (v. 2014‑07‑17)}.''
  \url{https://crawdad.org/roma/taxi/20140717/taxicabs}, 2014.

\bibitem{campbell}
L.~A. Urry, M.~L. Cain, S.~Wasserman, P.~Minorsky, and R.~Jane, {\em {Campbell
  Biology} (11th edition)}.
\newblock Pearson, 2017.

\bibitem{bacteriaapoptosis}
K.~W. Bayles, ``Bacterial programmed cell death: making sense of a paradox,''
  {\em Nature Reviews Microbiology}, vol.~12, no.~1, pp.~63--69, 2014.

\bibitem{fogmon2019}
A.~{Brogi}, S.~{Forti}, and M.~{Gaglianese}, ``{Measuring the Fog, Gently},''
  in {\em ICSOC}, pp.~523--538, 2019.

\bibitem{FORTI2020}
S.~Forti, M.~Gaglianese, and A.~Brogi, ``Lightweight self-organising
  distributed monitoring of {Fog} infrastructures,'' {\em Future Gener. Comput.
  Syst.}, vol.~114, pp.~605--618, 2021.
\newblock ({\textit{In press}}).

\bibitem{taherizadeh2018monitoring}
S.~Taherizadeh, A.~C. Jones, I.~Taylor, Z.~Zhao, and V.~Stankovski,
  ``Monitoring self-adaptive applications within edge computing frameworks: A
  state-of-the-art review,'' {\em J. Syst. Softw.}, vol.~136, pp.~19--38, 2018.

\bibitem{GuerreroLJ19}
C.~Guerrero, I.~Lera, and C.~Juiz, ``Evaluation and efficiency comparison of
  evolutionary algorithms for service placement optimization in fog
  architectures,'' {\em Future Gener. Comput. Syst.}, vol.~97, pp.~131--144,
  2019.

\bibitem{prologdebug}
W.~Drabent, ``The prolog debugger and declarative programming,'' in {\em
  International Symposium on Logic-Based Program Synthesis and Transformation},
  pp.~193--208, Springer, 2019.

\bibitem{brogi2019meet}
A.~Brogi, S.~Forti, C.~Guerrero, and I.~Lera, ``Meet genetic algorithms in
  monte carlo: optimised placement of multi-service applications in the fog,''
  in {\em 2019 IEEE International Conference on Edge Computing (EDGE)},
  pp.~13--17, IEEE, 2019.

\bibitem{pietri2016mapping}
I.~Pietri and R.~Sakellariou, ``Mapping virtual machines onto physical machines
  in cloud computing: A survey,'' {\em ACM Comput. Surv.}, vol.~49, no.~3,
  pp.~1--30, 2016.

\bibitem{TomarchioCM20}
O.~Tomarchio, D.~Calcaterra, and G.~D. Modica, ``Cloud resource orchestration
  in the multi-cloud landscape: a systematic review of existing frameworks,''
  {\em J. Cloud Comput.}, vol.~9, p.~49, 2020.

\bibitem{kadioglu2016heterogeneous}
S.~Kadioglu, M.~Colena, and S.~Sebbah, ``{Heterogeneous resource allocation in
  Cloud Management},'' in {\em NCA}, pp.~35--38, 2016.

\bibitem{yin2009rhizoma}
Q.~Yin, A.~Sch{\"u}pbach, J.~Cappos, A.~Baumann, and T.~Roscoe, ``Rhizoma: a
  runtime for self-deploying, self-managing overlays,'' in {\em Middleware
  2009}, pp.~184--204, 2009.

\bibitem{carlini2016self}
E.~Carlini, M.~Coppola, P.~Dazzi, M.~Mordacchini, and A.~Passarella,
  ``Self-optimising decentralised service placement in heterogeneous cloud
  federation,'' in {\em SASO}, pp.~110--119, 2016.

\bibitem{sathiaseelan2017towards}
A.~Sathiaseelan, M.~Selimi, C.~Molina, A.~Lertsinsrubtavee, L.~Navarro,
  F.~Freitag, F.~Ramos, and R.~Baig, ``Towards decentralised resilient
  community clouds,'' in {\em MECC}, pp.~1--6, 2017.

\bibitem{ferrer2019towards}
A.~J. Ferrer, J.~M. Marqu{\`e}s, and J.~Jorba, ``Towards the decentralised
  cloud: Survey on approaches and challenges for mobile, ad hoc, and edge
  computing,'' {\em ACM Comput. Surv.}, vol.~51, no.~6, pp.~1--36, 2019.

\bibitem{edgedeployment}
Z.~Xiang, S.~Deng, J.~Taheri, and A.~Y. Zomaya, ``Dynamical service deployment
  and replacement in resource-constrained edges,'' {\em Mob. Networks Appl.},
  vol.~25, no.~2, pp.~674--689, 2020.

\bibitem{COLISTRA201498}
G.~Colistra, V.~Pilloni, and L.~Atzori, ``The problem of task allocation in the
  internet of things and the consensus-based approach,'' {\em Comput.
  Networks}, vol.~73, pp.~98 -- 111, 2014.

\bibitem{greco16}
R.~Mennes, B.~Spinnewyn, S.~Latr{\'{e}}, and J.~F. Botero, ``{GRECO:} {A}
  distributed genetic algorithm for reliable application placement in hybrid
  clouds,'' in {\em CloudNet}, pp.~14--20, 2016.

\bibitem{guerrero2019lightweight}
C.~Guerrero, I.~Lera, and C.~Juiz, ``A lightweight decentralized service
  placement policy for performance optimization in fog computing,'' {\em J.
  Ambient Intell. Humaniz. Comput.}, vol.~10, no.~6, pp.~2435--2452, 2019.

\bibitem{bioinspired2020}
J.~{Herrera} and G.~{Moltó}, ``Toward bio-inspired auto-scaling algorithms: An
  elasticity approach for container orchestration platforms,'' {\em IEEE
  Access}, vol.~8, pp.~52139--52150, 2020.

\bibitem{ROSSI2020161}
F.~Rossi, V.~Cardellini, F.~{Lo Presti}, and M.~Nardelli, ``Geo-distributed
  efficient deployment of containers with kubernetes,'' {\em Comput. Commun.},
  vol.~159, pp.~161 -- 174, 2020.

\bibitem{hinrichs2009practical}
T.~L. Hinrichs, N.~S. Gude, M.~Casado, J.~C. Mitchell, and S.~Shenker,
  ``Practical declarative network management,'' in {\em WREN}, pp.~1--10, 2009.

\bibitem{herden2010declarative}
S.~Herden, A.~Zwanziger, and P.~Robinson, ``Declarative application deployment
  and change management,'' in {\em CNSM}, pp.~126--133, 2010.

\bibitem{edgeusher}
S.~Forti, F.~Paganelli, and A.~Brogi, ``{Probabilistic {QoS}-aware Placement of
  {VNF} chains at the Edge},'' {\em Theory and Practice of Logic Programming},
  2021.
\newblock {In press}.

\bibitem{secfog2019}
S.~{Forti}, G.~{Ferrari}, and A.~{Brogi}, ``{Secure Cloud-Edge Deployments,
  with Trust},'' {\em {Future Gener. Comput. Syst.}}, vol.~102, pp.~775--788,
  2020.

\bibitem{harzenetter2019automated}
L.~Harzenetter, U.~Breitenb{\"u}cher, F.~Leymann, K.~Saatkamp, B.~Weder, and
  M.~Wurster, ``Automated generation of management workflows for applications
  based on deployment models,'' in {\em EDOC}, pp.~216--225, 2019.

\bibitem{casadei19}
R.~Casadei and M.~Viroli, ``Coordinating computation at the edge: a
  decentralized, self-organizing, spatial approach,'' in {\em {FMEC} 2019},
  pp.~60--67, 2019.

\bibitem{casadei21}
D.~Pianini, R.~Casadei, M.~Viroli, and A.~Natali, ``Partitioned integration and
  coordination via the self-organising coordination regions pattern,'' {\em
  Future Gener. Comput. Syst.}, vol.~114, pp.~44--68, 2021.

\bibitem{aggregate}
M.~Viroli, J.~Beal, F.~Damiani, G.~Audrito, R.~Casadei, and D.~Pianini, ``From
  field-based coordination to aggregate computing,'' in {\em COORDINATION
  2018}, vol.~10852, pp.~252--279, Springer, 2018.

\bibitem{margariti2020modeling}
S.~V. Margariti, V.~V. Dimakopoulos, and G.~Tsoumanis, ``{Modeling and
  Simulation Tools for Fog Computing--A Comprehensive Survey from a Cost
  Perspective},'' {\em Future Internet}, vol.~12, no.~5, p.~89, 2020.

\bibitem{004}
H.~Gupta, A.~Vahid~Dastjerdi, S.~K. Ghosh, and R.~Buyya, ``{iFogSim: A toolkit
  for modeling and simulation of resource management techniques in the Internet
  of Things, Edge and Fog computing environments},'' {\em Soft. Pract. Exp.},
  vol.~47, no.~9, pp.~1275--1296, 2017.

\bibitem{sonmez2018edgecloudsim}
C.~Sonmez, A.~Ozgovde, and C.~Ersoy, ``{EdgeCloudSim}: An environment for
  performance evaluation of edge computing systems,'' {\em Trans. Emerg.
  Telecommun. Technol.}, vol.~29, no.~e3493, 2018.

\bibitem{fortipagiarobrogi2020}
S.~Forti, A.~Pagiaro, and A.~Brogi, ``{Simulating FogDirector Application
  Management},'' {\em Simul. Model. Pract. Theory}, vol.~101, no.~102021,
  pp.~1--18, 2020.

\bibitem{bacteriauav}
H.~A. Kurdi, M.~F. Aldaood, S.~Al{-}Megren, E.~Aloboud, A.~S. Aldawood, and
  K.~Youcef{-}Toumi, ``Adaptive task allocation for multi-uav systems based on
  bacteria foraging behaviour,'' {\em Appl. Soft Comput.}, vol.~83, 2019.

\bibitem{ahsan2020applications}
M.~M. Ahsan, K.~D. Gupta, A.~K. Nag, S.~Pouydal, A.~Z. Kouzani, and M.~P.
  Mahmud, ``Applications and evaluations of bio-inspired approaches in cloud
  security: A review,'' {\em IEEE Access}, 2020.

\bibitem{chang2018bacteria}
Y.-C. Chang, W.-X. Cai, and J.-W. Jhuang, ``Bacteria-inspired communication
  mechanism based on software-defined network,'' in {\em WOCC}, pp.~1--3, 2018.

\bibitem{bacteria5g}
H.~Chao, H.~Cho, T.~K. Shih, and C.~Chen, ``Bacteria-inspired network for 5g
  mobile communication,'' {\em {IEEE} Netw.}, vol.~33, no.~4, pp.~138--145,
  2019.

\end{thebibliography}

\end{document}